\documentclass[11pt,a4paper]{article}
\pdfoutput=1

\usepackage{jheppub}
\usepackage[T1]{fontenc}
\usepackage{extarrows}

\title{$U(1)$ current from the AdS/CFT: diffusion, conductivity and causality.}
\author[a]{Yanyan Bu,}
\author[a,b]{Michael Lublinsky,}
\author[a]{and Amir Sharon}
\affiliation[a]{Department of Physics, Ben-Gurion University of the Negev,
Beer-Sheva 84105, Israel}
\affiliation[b]{Physics Department, University of Connecticut, 2152 Hillside
Road, Storrs, CT 06269-3046, USA}

\emailAdd{yybu@post.bgu.ac.il}
\emailAdd{lublinm@bgu.ac.il}
\emailAdd{sharon.amir24@gmail.com}

\abstract{For a holographically defined finite temperature theory, we derive an off-shell  constitutive relation for a global $U(1)$ current driven by a weak external non-dynamical electromagnetic field. The constitutive relation involves an all order gradient expansion resummed into three momenta-dependent transport coefficient functions: diffusion, electric conductivity, and ``magnetic'' conductivity.  These transport functions are first computed analytically in the hydrodynamic limit, up to third order in the derivative expansion, and then numerically for generic values of momenta. We also compute a diffusion memory function, which, as a result of all order gradient resummation, is found to be causal.}


\arxivnumber{1511.08789}
\begin{document}
\maketitle

\flushbottom

\section{Introduction and Summary}\label{section1}
Hydrodynamics~\cite{fluid1,fluid2,km} is an effective long-distance description of most classical and quantum many-body systems at nonzero temperature. Within the hydrodynamic approximation, the entire dynamics of a microscopic theory is reduced to that of conserved macroscopic currents, such as expectation values of energy-momentum tensor or  charge current operators computed in a locally near equilibrium thermal state. An essential element of any hydrodynamics is a constitutive relation which relates the macroscopic currents to fluid-dynamic variables (fluid velocity, conserved charge densities, etc), and to external forces. Derivative expansion in the fluid-dynamic variables  accounts for deviations from thermal equilibrium.  At each order, the derivative expansion is fixed by thermodynamics and symmetries, up to a finite number of transport coefficients such as viscosity and diffusion coefficients. The latter are not calculable from hydrodynamics itself, but have to be determined from underlying microscopic theory or experimentally.

The most simple example of constitutive relation is the diffusion approximation for electric current
\begin{equation}\label{cr1}
\vec{J}=-\mathcal{D}_0\vec{\nabla}\rho
\end{equation}
where $\rho$ is the corresponding conserved charge density. Diffusion equation which follows from the current conservation has a well-known major conceptual problem---it violates causality. Relativistic Navier-Stokes hydrodynamics is plagued by similar pathology. In the case of the latter, this problem reveals itself in the form of numerical instabilities, rendering the framework into practically useless.

To restore causality one has to introduce higher order gradient terms in the constitutive relation. Moreover, inclusion of a finite number of terms is insufficient, and causality is restored only after all (infinite) order gradients are resummed. Generically, this is equivalent to a non-local constitutive relation of the type:
\begin{equation}\label{cr2}
\vec{J}(t)=-\int_{-\infty}^{\infty}dt^{\prime}\widetilde{\mathcal{D}}\left(t-t^\prime\right) \vec{\nabla} \rho(t^\prime)
\end{equation}
where $\widetilde{\mathcal{D}}$ is a memory function, which is most generally non-local  both in time and space. Causality implies that $\widetilde{\mathcal{D}}$ has no support  for $t<t^\prime$. In practice, the memory function is typically modelled: the simplest model imposes an exponential relaxation time approximation
$\widetilde{\mathcal{D}}(t-t^\prime) \sim e^{-(t-t^\prime)/\tau}\theta(t-t^\prime)$.
M$\ddot{\textrm{u}}$ller-Israel-Stewart~\cite{Muller,Israel,IS1976,IS1979} second order relativistic hydrodynamics, commonly used in simulations of heavy ion collisions, is an example of such a modelling\footnote{See~\cite{1404.4894} for a baryon diffusion model which addresses the causality issue.}.

AdS/CFT correspondence~\cite{hep-th/9711200,hep-th/9802109,hep-th/9802150} opens a possibility to study transport properties exactly and partially analytically, at least for a class of quantum gauge theories for which gravity duals can be constructed. This holographic duality maps hydrodynamic fluctuations of a boundary fluid into long-wavelength gravitational perturbations of a stationary black brane in asymptotic AdS space~\cite{hep-th/0104066,hep-th/0205052,hep-th/0210220}. Viscosity and many other transport coefficients could be computed from the gravity side of the correspondence. The ratio of viscosity $\eta_{_{0}}$ to the entropy density $s$ was computed in~\cite{hep-th/0104066,hep-th/0205052,hep-th/0405231}
\begin{equation}\label{ratio}
\frac{\eta_{_{0}}}{s}=\frac{1}{4\pi}
\end{equation}
and was found to be universal for all gauge theories with Einstein gravity duals~\cite{hep-th/0311175,0809.3808,0808.3498}. Remarkably, the fluid/gravity correspondence is not limited to linear response theory for small perturbations of the velocity field $u_\mu$. Navier-Stokes equations are completely encoded in the Einstein equations. Particularly, the formalism of~\cite{0712.2456} provides a systematic framework to construct nonlinear fluid dynamics, order by order in the fluid velocity derivative expansion, with the transport coefficients determined from the gravity side. The study of~\cite{0712.2456} was subsequently generalised to conformal fluids in higher dimensions~\cite{0806.4602}, weakly curved background manifolds~\cite{0809.4272}, and to forced fluids~\cite{0806.0006}. We refer the reader to~\cite{0704.0240,0905.4352,1107.5780} for comprehensive reviews of fluid/gravity correspondence.

In~\cite{1406.7222,1409.3095}, two of us built upon previous work~\cite{0905.4069} and constructed a flat space \emph{all-order linearly resummed} relativistic conformal hydrodynamics using the fluid/gravity correspondence. By {\it linear} in the fluid dynamic variables, we mean small amplitude terms like $(\nabla\nabla\cdots\nabla u)_{\mu\nu}$; but with all nonlinear structures such as $(\nabla u)^2_{\mu\nu}$ neglected. This is mathematically a well controlled approximation. The relativistic hydrodynamics with all order derivatives resummed was found to have a rich structure, absent in a strict low frequency/momentum approximation. The fluid stress-energy tensor was expressed using the shear term  with $\eta_0$ replaced by a viscosity functional $\eta$ of space-time derivative operators, and a new viscous term, which emerged starting from the third order in the gradient expansion. In Fourier space, both transport coefficient functionals became functions of frequency and spatial momentum, $\eta(\partial_t,\vec \partial^2)\rightarrow \eta(\omega, q^2)$. We performed a similar study for the Einstein-Gauss-Bonnet gravity in~\cite{1504.01370}. Ref.~\cite{1502.08044} generalised computations of~\cite{1406.7222,1409.3095} by  including small non-dynamical metric perturbations in the boundary stress tensor. It was found that four new transport coefficient functions are needed in order to linearly couple the stress tensor to external gravitational perturbations. In~\cite{0905.4069} these transport coefficient functions were called gravitational susceptibilities of the fluid (GSF). They play a role similar to electrical conductivity, which will be discussed below. After the resummation, the memory function associated with the viscosity term was found to be consistent with the causality constraints.

Our strategy in~\cite{1406.7222,1409.3095,1502.08044,1504.01370} was to further develop the method of~\cite{0712.2456, 0806.0006} by including all order linear structures in a self-consistent manner. Interestingly, we found that the dynamical components of the bulk Einstein equations are sufficient to compute an ``off-shell'' stress-energy tensor of the boundary fluid, with the transport coefficient functions fully determined. Additionally imposing four remaining constraints from the set of all bulk equations is equivalent to putting ``on-shell'' thus obtained stress-energy tensor.

In the present work, we extend our study into dynamics of a conserved vector current associated with a global $U(1)$ symmetry, focusing on the properties of all order charge diffusion and conductivity. The most general linear in the fields (small amplitude) covariant form constitutive relation extending~(\ref{cr1},\ref{cr2}) is\footnote{A very similar constitutive relation was deduced from an effective action approach in the most recent paper~\cite{1511.03646}, where the chemical potential (instead of the charge density $\rho$) was used to express $J^\mu$. However, no values for the transport coefficients were determined there.}
\begin{equation}\label{constitutive relation-cov}
J^{\mu}(x_\sigma)=\rho(x_\sigma)\, u^{\mu}-\mathcal{D}P^{\mu\nu}\partial_\nu \rho(x_\sigma) +\sigma_e u^\alpha P^{\mu\nu}F_{\alpha\nu}^{(0)}(x_\sigma)+\sigma_m P^{\alpha\beta}P^{\mu\nu}\partial_\alpha F_{\beta\nu}^{(0)}(x_\sigma),
\end{equation}
where $u^{\alpha}=(1,0,0,0)$ and $P^{\mu\nu}=\eta^{\mu\nu}+u^\mu u^\nu$ is a projection operator onto spatial directions. The field strength tensor  $F_{\mu\nu}^{(0)}=\partial_\mu A_\nu^{(0)}-\partial_\nu A_\mu^{(0)}$ is associated with a non-dynamical external electromagnetic potential $A_\mu^{(0)}$ (see equation~(\ref{ansatz})). More transparently, the constitutive relation~(\ref{constitutive relation-cov}) reads
%
%
\begin{equation}\label{constitutive relation}
J^t=\rho,~~~~~~~~~\vec{J}=-\mathcal{D}\vec{\nabla}\rho+\sigma_e \vec{E}+\sigma_m\vec{\nabla} \times \vec{B}
\end{equation}
The transport coefficients $\mathcal{D}$, $\sigma_e$ and $\sigma_m$ are scalar functionals of spacetime derivatives
\begin{equation}
\mathcal{D}\left[\partial_t,\vec{\partial}^2\right],~~~ \sigma_e\left[\partial_t,\vec{\partial}^2\right],~~~
\sigma_m\left[\partial_t,\vec{\partial}^2\right].
\end{equation}
In the Fourier space, via  replacement $(\partial_t,\vec{\partial}^2)\longrightarrow(-i\omega,-{q}^2)$, these functionals of the derivative operators turn into complex
functions of frequency $\omega$ and momentum $q$:
\begin{equation}\label{constitutive momenta}
\vec{J}(\omega,\vec{q})=-\mathcal{D}\left(\omega,q^2\right)i\vec{q}\;\rho(\omega,\vec{q}) +\sigma_e\left(\omega,{q}^2\right)\vec{E}(\omega,\vec{q})+\sigma_m\left(\omega,
{q}^2 \right) i\vec{q}\times \vec{B}\left(\omega,\vec{q}\right).
\end{equation}

In the hydrodynamic limit, the transport coefficients are expandable in  Taylor series, each term corresponding to a fixed order gradient
\begin{eqnarray}
&&\mathcal{D}(\omega\rightarrow 0,q^2\rightarrow 0)=\mathcal{D}_0+ i\tau_{_D} \omega  + \lambda^1_{_D}\omega^2+\lambda^2_{_D}q^2+\cdots, \\
&&\sigma_{e,m}(\omega\rightarrow 0,q^2\rightarrow 0)=\sigma_{e,m}^0+ i\tau_{e,m} \omega  + \sigma^1_{e,m}\omega^2+\sigma^2_{e,m}q^2+\cdots.
\end{eqnarray}
$\mathcal{D}$ generalises the usual concept of diffusion constant $\mathcal{D}_0$ and will be referred to as diffusion function. $\sigma_e$ is an extension of electric DC conductivity $\sigma_{e}^0$. The second conductivity $\sigma_m$ reflects a response to external magnetic field. This second derivative structure (the curl of magnetic field) appeared already in~\cite{1003.0699,1105.6360} but no information about $\sigma_m$ was provided there. $\tau_{_D,e,m}$ are corresponding relaxation times. The constitutive relation~(\ref{constitutive relation}) is an off-shell construction. Nevertheless, for a specific underlying microscopic theory, we are able to determine these transport coefficient functions. Putting the current on-shell reproduces the usual linear response theory, such as the Ohm's law\footnote{It might be reasonable to call $\sigma_{e,m}$ ``off-shell conductivities'' as they differ from the ``on-shell'' AC conductivity entering the Ohm's law.}. While the magnetic term drops from the continuity equation
\begin{equation}\label{continuity}
\partial_\mu J^\mu=0,
\end{equation}
$\sigma_m$ does contribute to current-current two-point correlators (see~(\ref{correlators})). Our construction is quite analogous to that of all-order stress-energy tensor derived in~\cite{1502.08044}: $\mathcal{D}$ is analogous to the viscosity $\eta$,  while $\sigma_e,\sigma_m$ play the role similar to the GSFs in the constitutive relation for the stress tensor.

Our goal below is to compute $\mathcal{D}$, $\sigma_e$, and $\sigma_m$ for a specific 4d theory defined holographically: the bulk theory is a \emph{probe} Maxwell field in the \emph{background} of Schwarzschild-$AdS_5$ geometry. This theory is a consistent reduction  from probe D-brane models such as D3/D7 construction of~\cite{hep-th/0205236}. In  quenched approximation, i.e., when the number of probe D-branes is much smaller than that of colour branes responsible for the bulk geometry, it is reasonable to neglect the back-reaction of the bulk Maxwell field on the metric. Thus we study charge dynamics decoupled from any energy-momentum flow. Coupling the dynamics of the $U(1)$ current to that of the stress tensor is certainly interesting and is likely to reveal new transport structures. We plan to explore this in the future.

$R$-current of the boundary CFT is a particular example of the $U(1)$ current we are interested in. Its hydrodynamics was previously studied via holography in~\cite{hep-th/0205052,hep-th/0210126,hep-th/0309213,hep-th/0607237,hep-th/0701036, 0805.2570,0806.0110,0806.4460,0810.1077,0706.0162,0903.2834}. For various supergravity backgrounds, ref.~\cite{hep-th/0309213} derived a simple expression for the diffusion constant $\mathcal{D}_0$ in terms of the metric components. In~\cite{0809.3808,1006.1902,1109.2698}, black hole membrane paradigm was used to explore universality of the diffusion constant and DC conductivity. For $\mathcal{N}=4$ $SU(N_c)$ SYM plasma in the large $N_c$ and large 't Hooft coupling limit, the $R$-charge diffusion constant $\mathcal{D}_0=1/(2\pi T)$, DC conductivity $\sigma_e^0=\pi T$ and $\tau_e=\log 2/(2\pi T)$. For convenience, below we will adopt dimensionless $\omega$ and $q^2$ normalised to $\pi$ times the temperature, $\pi T=1$. The physical frequency and spatial momentum are $\pi T \omega$ and $\pi T {q}$.

It is important to stress that the holographically defined boundary theory does not have a dynamical $U(1)$ gauge field. Hence the background external fields ($\vec{E}$ and $\vec{B}$) are by construction non-dynamical. However, one could imagine gauging this global $U(1)$ symmetry~\cite{hep-th/0607237,0806.0110} with a small coupling.
Then  the current $J^\mu$ can be treated as an induced current in a conducting medium described by a self-consistent macroscopic electrodynamics (see section~\ref{subsection44}).

Our derivation consists of several steps. We first solve the bulk Maxwell equations in a static $AdS_5$ black brane geometry.  An important element of the procedure is that we solve the Maxwell equations for a set of unspecified boundary conditions $A_\mu^{(0)}$ imposed at the conformal boundary. The boundary field $A_\mu^{(0)}$ acts as a source of the current $J_\mu$. All order transport coefficient functions are then read off from the near boundary behaviour of the bulk Maxwell field. In the hydrodynamic limit $\omega \ll 1, q\ll 1$, our analytical results derived in subsection~\ref{subsection51} are
\begin{equation}\label{hydro expansion}
\begin{split}
\mathcal{D}&=\frac{1}{2}+\underline{\frac{1}{8}\pi i\omega+\frac{1}{48}\left[-\pi^2 \omega^2+ q^2 \left(6\log 2-3\pi\right)\right]}+\cdots,\\
\sigma_e&=1+\frac{\log 2}{2}i\omega+\frac{1}{24}\left[\pi^2\omega^2\underline{-q^2\left(3\pi+ 6\log 2 \right)} \right]+\cdots,\\
\sigma_m&=0+\underline{\frac{1}{16}i\omega\left(2\pi-\pi^2+4\log 2\right)}+\cdots.
\end{split}
\end{equation}
These results fully agree with  all previous studies (and extend beyond): the constant term in $\mathcal{D}$ ($\mathcal{D}_0$) was first reported in~\cite{hep-th/0205052}; the DC conductivity $\sigma_e^0$ is consistent with that of~\cite{hep-th/0607237} obtained
using the low-frequency results of~\cite{hep-th/0205052}; the expression for $\sigma_e$ at $q^2=0$ is in agreement with~\cite{0810.1077,0706.0162} (see also subsection~\ref{subsection52} for a more comprehensive comparison extending to generic values of $\omega$). The underlined terms are new results unavailable in the literature. Quite interestingly, the hydrodynamic expansion of $\sigma_m$ starts from a term linear in $\omega$, rather than from a constant.

For generic values of momenta, the transport coefficient functions are computed numerically. The results are presented and discussed in subsection~\ref{subsection52}.
Particularly, the diffusion function vanishes at very large momenta, which is a necessary condition for causality restoration. The memory function introduced in~(\ref{cr2}) is defined as Fourier transform of the diffusion function $\mathcal{D}$
\begin{equation}\label{memory fun1}
\widetilde{\mathcal{D}}\left(t,q^2\right)=\frac{1}{\sqrt{2\pi}}\int_{-\infty}^\infty \mathcal{D}\left(\omega, q^2\right)e^{-i\omega t}d\omega.
\end{equation}
$\widetilde{\mathcal{D}}$ is also computed in subsection~\ref{subsection52} and it is found to be causal $\widetilde{\mathcal{D}}(t)\sim \Theta(t)$. For positive times, $\widetilde{\mathcal{D}}(t)$ displays damped oscillations, which reflects the presence of a (infinite) set of complex poles---the bulk's quasi-normal modes. In the next section~\ref{section2}, the meaning of all-order gradient resummation and its implication on the dynamics are briefly clarified, essentially quoting the discussion presented earlier in~\cite{1502.08044}.

The holographic model is introduced in section~\ref{section3}. Section~\ref{section4} presents our main calculations. Maxwell equations in the bulk are discussed first. We search for solutions which are functionals of the boundary data $\rho(x_\alpha)$ and $A_\mu^{(0)}(x_\alpha)$. Decomposing these solutions in a basis constructed from the boundary fields, we are able to cast the problem into a set of \emph{ordinary} differential equations for the decomposition coefficients. Then, we read off $J^\mu$ from the near-boundary expansion of the decomposition coefficients. The latter are computed  first analytically and then numerically in section~\ref{subsection42}. In subsection~\ref{subsection43} we derive the continuity equation from the constraint component of the bulk Maxwell equation. Current-current two-point correlators are also expressed in terms of the transport coefficients $\mathcal{D}$, $\sigma_{e,m}$.

In section~\ref{subsection44}, we gauge the external field, turning it into a dynamical Maxwell field on the boundary. Combined with the current constitutive
relation~(\ref{constitutive relation}) this becomes a self-consistent electrodynamics of a conducting medium.  For completeness of our presentation
we compute the medium's dielectric functions (transverse and longitudinal) and  remark on negative refraction phenomena. In section~\ref{summary} we make a brief summary.
The calculations presented in section~\ref{section4} are derived in Landau frame. In Appendix~\ref{appLF} we demonstrate that in fact the results are frame independent.
Perturbative solutions for the bulk Maxwell fields are given in Appendix~\ref{appendix}.

\section{All order gradient resummation}\label{section2}

In this section we provide a clarification about what we actually mean by all-order gradient resummation, which along spacial derivatives includes an infinite number of time derivatives. This particularly means that the effective dynamical equations  require an infinite set of initial conditions or, equivalently, the dynamics at hand is a theory of infinite number of degrees of freedom. Indeed, these are the quasi-normal modes of the bulk Maxwell theory.  An alternative way to see the origin of the problem is to realise that exact dynamics in the bulk cannot be mapped onto a single degree of freedom on the boundary.

Yet, it turns out that it is possible to formulate this all-order hydrodynamics as a normal initial value problem with all the time derivatives absorbed into the memory function. For the sake of argument we will ignore any space dependence and will set $B=0$. The causal current
\begin{equation}
\vec J(t)= \int_{-\infty}^t dt^\prime \left[\widetilde D(t-t^\prime) \vec\nabla \rho(t^\prime)\,+\,\tilde \sigma_e(t-t^\prime)\, \vec E(t^\prime)\right].
\end{equation}
and $\tilde \sigma_e$ is a memory  associated with the conductivity function.
The external electric field $E$ is normally turned on at negative times, so  to create an  initial charge density profile at $t=0$, and then turned off at $t=0$ ($E(t>0)=0$), letting the system to freely relax to its equilibrium at infinite future. For such experimental setup, for positive times the current
\begin{equation}\label{J2}
\vec J(t>0)= \int_{0}^t dt^\prime \widetilde D(t-t^\prime) \vec\nabla \rho(t^\prime)\,+\,\vec J_H(t)
 \end{equation}
with
\begin{equation}\label{JH}
\vec J_H(t)= \int_{-\infty}^0 dt^\prime \left[ \widetilde D(t-t^\prime) \vec\nabla \rho(t^\prime)\,+\,\tilde \sigma_e(t-t^\prime)\,\vec E(t^\prime)\right].
\end{equation}
Generically, $J_H$ is not vanishing and it accounts for the entire history of the system at negative times. This is in contrast to a typical memory function-based approach, where one introduces constitutive relation~(\ref{J2}) assuming $J_H=0$ and then also models
$\widetilde D$ \cite{km,fluid2}.

Our construction is so far formally exact. However, in order to solve the dynamical
equation (\ref{continuity}), it is not sufficient to provide the initial condition for the density only, but we also need the ``history'' current $J_H$ at all times,
equivalent to providing infinitely many additional initial conditions.

We are now to discuss under what conditions we can nevertheless set $J_H$ to zero, casting our theory into a well-defined initial value problem. The response functions $\widetilde D$ and $\tilde \sigma_e$ are some given functions defined by underlying microscopic theory. Thus the equation $J_H(t>0)=0$ is in fact an equation for the electric field $E$ at $t<0$. It is, however, not obvious that there exists a solution for generic $\widetilde D$ and $\tilde \sigma$, because we want the current $J_H=0$ at all positive times. Even though we cannot guarantee vanishing of the current identically, it is safe to assume that its effect could be rendered negligibly small. Particularly, it is obvious that $J_H$ vanishes at late times and its only potential influence could be at very early times when the current $J\simeq J_H$.

\section{The holographic model}\label{section3}

We would like to study a finite charge density/chemical potential system having a global $U(1)$ symmetry exposed to an external electromagnetic field.
The holographic model is a Maxwell field in the \emph{background} of Schwarzschild-$AdS_5$ geometry.


{\bf Conventions}:  the upper case Latin indices $\left\{M,N,\cdots\right\}$  denote the bulk while the lower case Greek indices $\left\{\mu,\nu,\cdots\right\}$ the boundary coordinates; the lower case Latin indices $\left\{i,j,\cdots\right\}$ indicate the spatial directions on the boundary; $g_{_{MN}}$ is the bulk metric; $\gamma_{\mu\nu}$ is the  induced metric on a hyper-surface $\Sigma$  of constant radial coordinate $r$, and $\eta_{\mu\nu}$ is a Minkowski metric at the boundary. The gauge coupling of the bulk Maxwell field is set to unity.

The regularised bulk action is
\begin{equation}\label{maxwell action}
S=\frac{1}{4}\int d^5x\sqrt{-g}F^{MN} F_{MN} + S_{\textrm{c.t.}},
\end{equation}
where $F_{MN}=\nabla_{M}A_N- \nabla_{N}A_M$, and $S_{\textrm{c.t.}}$ is a counter-term action to be specified below.
From the action~(\ref{maxwell action}), the Maxwell equations are
\begin{equation}\label{Maxwell eq}
\textrm{EQ}^N:=\nabla_{M}F^{MN}=0\Longrightarrow\partial_M\left(\sqrt{-g}F^{MN}\right)=0.
\end{equation}
Radial gauge $A_r=0$ is used throughout this work. In the ingoing Eddington-Finkelstein coordinates, the Schwarzschild-$AdS_5$ geometry is
\begin{equation}\label{AdS5}
ds^2=g_{_{MN}}dx^Mdx^N=2dtdr-r^2f(r)dt^2+r^2\delta_{ij}dx^idx^j,
\end{equation}
where $f(r)=1-1/r^4$. The horizon is set to be at $r=1$, which normalises the Hawking temperature to $\pi T=1$.

Holographic renormalisation of the $U(1)$ gauge field in the asymptotically $AdS_5$ space was  considered previously using the Fefferman-Graham  coordinates~\cite{hep-th/0002125,hep-th/0112119,1004.3541}. We are going to revise these studies for the case of  the Eddington-Finkelstein coordinates.
First, the Maxwell equations~(\ref{Maxwell eq}) are assumed to be solved.  Near the conformal boundary $r=\infty$ the solution is expandable in a series
\begin{equation}\label{boundary solution}
A_{\mu}(r,x_{\alpha})= A_{\mu}^{(0)}(x_{\alpha})+\frac{A^{(1)}_{\mu} (x_{\alpha})}{r} +\frac{A_{\mu}^{(2)}(x_{\alpha})}{r^2} + \frac{B_{\mu}^{(2)} (x_{\alpha})} {r^2}\log r^{-2}+\mathcal{O}\left(\frac{\log r^{-2}}{r^3}\right),
\end{equation}
where $A_{\mu}^{(1)}$ and $B_{\mu}^{(2)}$ are fixed by the Maxwell equations~(\ref{Maxwell eq})
\begin{equation}\label{subleading terms}
A_{\mu}^{(1)}= F_{t\mu}^{(0)},~~~~4B_{\mu}^{(2)}=\partial^{\nu}F^{(0)}_{\mu\nu}.
\end{equation}
The values of $A_\mu^{(0)}$ and $A_\mu^{(2)}$ are left unspecified. These are  the equation's integration constants.
Second, the near-boundary expansion for $A_M$~(\ref{boundary solution}) is substituted into the Maxwell action of~(\ref{maxwell action}), resulting in a logarithmical divergence near the conformal boundary $r=\infty$,
\begin{equation}
S_{\textrm{reg.}}=\frac{1}{4}\int_1^{\Lambda \to \infty} dr \int d^4x \sqrt{-g} F^{MN}F_{MN}= \frac{1}{4} \log \Lambda \int d^4x F^{(0)}_{\mu\nu}F^{\mu\nu(0)}+\cdots
\end{equation}
where ``$\cdots$'' denotes a finite piece when  $\Lambda\to \infty$.
$S_{\textrm{reg.}}$ is a regularised action at the cutoff $\Lambda$. Its covariant form
makes up  the  counter-term  in~(\ref{maxwell action})
\begin{equation}\label{counter term}
S_{\textrm{c.t.}}=-\frac{1}{4}\log r\,\int d^4x\sqrt{-\gamma}F_{\mu\nu} F^{\mu\nu},
\end{equation}
where the indices are contracted with the  metric $\gamma_{\mu\nu}$ on $\Sigma$:
\begin{equation}
ds^2\mid_{\Sigma}=\gamma_{\mu\nu}dx^\mu dx^\nu=-r^2f(r)dt^2+r^2\delta_{ij}dx^idx^j
\end{equation}
In writing down the counter-term~(\ref{counter term}) only the minimal subtraction scheme is implemented, that is (\ref{counter term}) has no finite piece. Keeping some
in the counter-term  would induce certain contact terms (see e.g.~\cite{0902.0409,1206.6785} for discussion of  ambiguities in the renormalisation scheme).

The boundary current is defined as a variation  of the on-shell action $S$ with respect to the source:
\begin{equation}
J^{\mu}\equiv \lim_{r\to\infty}\frac{\delta S}{\delta A_{\mu}}.
\end{equation}
The variation of $S$ is computed using equations of motion~(\ref{Maxwell eq}) and assuming $A_\mu$ falls off sufficiently quickly at infinity
\begin{equation}\label{variation}
\begin{split}
\delta S
&=\int d^5x\sqrt{-g}\nabla_M \left(\delta A_N F^{MN}\right)- \log r \int d^4x\sqrt{-\gamma}\left[\widetilde{\nabla}_\mu \left(\delta A_\nu F^{\mu\nu}\right) - \delta A_\nu \widetilde{\nabla}_\mu F^{\mu\nu}\right]\\
&=\int d^4x \sqrt{-\gamma} \left\{\delta A_N F^{MN} n_{_M}+  \delta A_\mu \widetilde{\nabla}_\nu F^{\nu\mu}\log r\right\},
\end{split}
\end{equation}
where $\widetilde{\nabla}_{\mu}$ is compatible with
 $\gamma_{\mu\nu}$, and $n_{_M}$ is the unit out-pointing vector normal to  $\Sigma$,
\begin{equation}
n_{_M}\equiv \frac{\nabla_M r}{\sqrt{g^{AB}\nabla_{A} r \nabla_{B} r}}.
\end{equation}
The boundary current is
\begin{equation}
J^{\mu}=\lim_{r\to\infty} \left\{ \sqrt{-\gamma} F^{M\mu}n_{_M} +\sqrt{-\gamma}\widetilde{\nabla}_{\nu}F^{\nu\mu}\log r \right\}.
\end{equation}
Expressed in terms of the near-boundary asymptotic expansion, (\ref{boundary solution}) becomes
\begin{equation}\label{jmu formula2}
J^{\mu}=-\eta^{\mu\nu}\left(2A_{\nu}^{(2)}+2B_{\nu}^{(2)}+\eta^{\sigma t}\partial_\sigma F_{t\nu}^{(0)}\right).
\end{equation}

\section{Boundary current from the Maxwell dynamics in the bulk}\label{section4}

The five Maxwell equations~(\ref{Maxwell eq}) split into  four dynamical components ${\textrm{EQ}}^\mu=0$ and one constraint ${\textrm{EQ}}^r=0$.
Following the strategy developed in~\cite{1406.7222,1409.3095,1502.08044,1504.01370},
we will initially solve the dynamical equations only. Substituting the solution into~(\ref{jmu formula2})
will result in an ``off-shell''  current $J^\mu$ with values of transport coefficients being fully determined.
The constraint ${\textrm{EQ}}^r=0$ is then found to be equivalent to the continuity equation~(\ref{continuity}).

We first notice that the equations ${\textrm{EQ}}^\mu=0$ admit the most general static homogeneous solutions
\begin{equation}\label{static}
A_\mu=A_\mu^{(0)}+\frac{\rho}{2r^2}\delta_{\mu t},
\end{equation}
where $A_\mu^{(0)}$ and $\rho$ are, for the moment, assumed to be some constants.
Substituting~(\ref{static}) into (\ref{jmu formula2}), the boundary current corresponding to this solution  is just
\begin{equation}
J^t=\rho, \ \ \ \ \ \ \ \ J^i=0
\end{equation}
Hence we identify $\rho$ with the expectation value of the charge density operator.  The boundary theory is a static uniformly charged plasma with no
external fields present ($F_{\mu\nu}^{(0)}=0$).

Next, following the spirit of~\cite{0712.2456},
we promote the constant parameters $A_\mu^{(0)}$ and $\rho$ into unspecified  functions of the boundary coordinates,
\begin{equation}
A_\mu^{(0)}\to A_\mu^{(0)}(x_\alpha),~~~~~\rho\to\rho(x_\alpha).
\end{equation}
$A_\mu$ of (\ref{static}) would cease to solve the Maxwell equations for arbitrary functions $A_\mu^{(0)}(x_\alpha)$ and $\rho({x_\alpha})$.
The solution  has to be amended:
\begin{equation}\label{ansatz}
A_{\mu}(r,x_{\alpha})=A_{\mu}^{(0)}(x_{\alpha})+\frac{\rho(x_{\alpha})}{2r^2}\delta_{\mu t}+a_{\mu}(r,x_{\alpha}).
\end{equation}
%
%
For given $A_\mu^{(0)}(x_\alpha)$ and $\rho({x_\alpha})$,  the Maxwell equations~(\ref{Maxwell eq}) will be solved for
the correction $a_\mu$. The equations are subject to specific boundary conditions to be discussed next.
Near $r=\infty$, we require that
\begin{equation}\label{AdS constraint1}
a_{\mu}\left(r\to\infty,x_\alpha\right)\rightarrow 0,
\end{equation}
%
%
so that $A_\mu^{(0)}(x_\alpha)$ could be identified with the external electromagnetic field sourcing  the  current $J^\mu$.
We would like to keep identifying $\rho(x_\alpha)$ as the charge density profile, equivalent to
Landau frame convention
\begin{equation}\label{landau frame}
J^\mu u_\mu=-\rho(x_\alpha).
\end{equation}
The Landau frame convention can be argued to be equivalent to residual gauge fixing. In Appendix~\ref{appLF} we relax this condition and demonstrate that the results on all the transport coefficients are still uniquely fixed\footnote{We thank Hong Liu for raising up this question.}. Finally, since we are working in the ingoing Eddington-Finkelstein coordinate, we further require that $a_\mu(r,x_\alpha)$ is regular over the whole range of $r$. Particularly,
the regularity requirement at the horizon $r=1$ is equivalent to the in-falling wave condition of~\cite{hep-th/0205051} in the Poincare patch.
The above boundary conditions are sufficient to fix $a_\mu$ uniquely.

The Maxwell equations~(\ref{Maxwell eq}) reduce to the following equations for $a_\mu$,
\begin{equation}\label{eq_a}
\begin{split}
&\textrm{EQ}^t=0\Rightarrow 0=r^2\partial_r^2a_t+3r\partial_ra_t+\partial_r \partial_k a_k,\\
&\textrm{EQ}^i=0\Rightarrow 0=(r^5-r)\partial_r^2a_i+(3r^4+1)\partial_ra_i+2r^3\partial_r \partial_ta_i -r^3\partial_r \partial_i a_t,\\
&~~~~~~~~~~~~~~~~~~~~+r^2(\partial_ta_i-\partial_ia_t)+r(\partial^2a_i-\partial_i\partial_k a_k)+ \frac{1}{2} \partial_i\rho 
\\
&~~~~~~~~~~~~~~~~~~~~
+r^2\left(\partial_t A_i^{(0)} -\partial_i A_t^{(0)}\right) +r\left(\partial^2A_i^{(0)}- \partial_i\partial_k A_k^{(0)} \right),
\end{split}
\end{equation}
where the inhomogeneous terms are induced by the fields $A_{\mu}^{(0)}(x_\alpha)$ and $\rho(x_{\alpha})$. Thus, the solutions
$a_\mu$ are linear functionals of these fields. Following  a similar procedure in~\cite{1406.7222,1409.3095,1502.08044,1504.01370}, $a_{\mu}$ are decomposed in a basis made of $A_{\mu}^{(0)}(x_\alpha)$ and $\rho(x_{\alpha})$ fields
\begin{equation}\label{basis}
\begin{split}
a_t&=S_1A_t^{(0)}+S_2\partial_kA_k^{(0)}+S_3\rho,\\
a_i&=V_1A_i^{(0)}+V_2\partial_iA_t^{(0)}+V_3\partial_i\partial_kA_k^{(0)}+V_4\partial_i \rho,
\end{split}
\end{equation}
The decomposition coefficients $S_i$ and $V_i$ are functions of $r$ and also  functionals of the boundary derivative operators $\partial_t$ and $\partial^2_i$. In Fourier space, $S_i$ and $V_i$ become functions of $(r,\omega, q^2)$. The equations~(\ref{eq_a}) for $a_\mu$ map into a system of second order {\em ordinary} differential equations. We summarised these equations by grouping them into decoupled sub-sectors.
\paragraph{I:~$\left\{S_1,V_2\right\}$}
\begin{equation}\label{S1V2}
\left\{
\begin{aligned}
0=&r^2\partial_r^2S_1+3r\partial_rS_1-q^2\partial_rV_2,\\
0=&(r^5-r)\partial_r^2V_2+(3r^4+1)\partial_rV_2-2i\omega r^3\partial_rV_2\\
&-i\omega r^2V_2-r^3\partial_rS_1-r^2S_1-r^2
.
\end{aligned}
\right.
\end{equation}
\paragraph{II:~$\left\{S_2,V_1,V_3\right\}$}
\begin{equation}\label{S2V1V3}
\left\{
\begin{aligned}
0=&r^2\partial_r^2S_2+3r\partial_rS_2+\partial_rV_1-q^2\partial_rV_3,\\
0=&(r^5-r)\partial_r^2V_1+(3r^4+1)\partial_rV_1-2i\omega r^3\partial_rV_1\\
&-i\omega r^2V_1-q^2rV_1-i\omega r^2-q^2r,\\
0=&(r^5-r)\partial_r^2V_3+(3r^4+1)\partial_rV_3-2i\omega r^3\partial_r V_3\\
&-i\omega r^2V_3-r^3\partial_rS_2-r^2S_2-rV_1-r
.
\end{aligned}
\right.
\end{equation}
\paragraph{III:~$\left\{S_3,V_4\right\}$}
\begin{equation}\label{S3V4}
\left\{
\begin{aligned}
0=&r^2\partial_r^2S_3+3r\partial_rS_3-q^2\partial_rV_4,\\
0=&(r^5-r)\partial_r^2V_4+(3r^4+1)\partial_rV_4-2i\omega r^3\partial_rV_4\\
&-i\omega r^2V_4-r^3\partial_rS_3-r^2S_3+\frac{1}{2}.
\end{aligned}
\right.
\end{equation}
The boundary conditions for $S_i$ and $V_i$  are derivable from those for $a_\mu$.
The condition (\ref{AdS constraint1}) translates into
\begin{equation}\label{AdS constraint}
S_i\longrightarrow 0,~~~V_i\longrightarrow 0~~\textrm{as}~~r\to\infty.
\end{equation}
The regularity condition for $a_\mu(r)$ is equivalent to that  for the coefficients $S_i$ and $V_i$.
Finally, the Landau frame choice becomes a non-trivial constraint on the pre-asymptotic behaviour of the coefficients $S_i$ (see below).


Since the boundary current $J_\mu$ (\ref{jmu formula2}) is defined in terms of the near-boundary behaviour of the field $A_\mu$,
our next step is to re-express it in terms of the pre-asymptotics  of the coefficients $S_i$ and $V_i$.
The coefficients $S_i$ and $V_i$  near the boundary $r=\infty$ have the form
\begin{equation}\label{boundary behavior}
\begin{split}
S_1&\longrightarrow \frac{s_{_1}(\omega,q^2)}{r^2}-\frac{q^2\log r}{2r^2}+\mathcal{O} \left(\frac{\log r}{r^3}\right),\\
S_2&\longrightarrow \frac{s_{_2}(\omega,q^2)}{r^2}+ \frac{i\omega \log r}{2r^2}+ \mathcal{O} \left(\frac{\log r}{r^3} \right)\\
S_3&\longrightarrow\frac{s_{_3}(\omega,q^2)}{r^2}+\mathcal{O}\left(\frac{1}{r^3}\right),\\
V_1&\longrightarrow-\frac{i\omega}{r}+\frac{v_{_1}(\omega,q^2)}{r^2}+\frac{1}{2}(\omega^2-q^2) \frac{\log r}{r^2}+\mathcal{O}\left(\frac{\log r}{r^3}\right),\\
V_2&\longrightarrow-\frac{1}{r}+\frac{v_{_2}(\omega,q^2)}{r^2}-\frac{i\omega\log r}{2r^2}+ \mathcal{O}\left(\frac{\log r}{r^3}\right),\\
V_3&\longrightarrow\frac{v_{_3}(\omega,q^2)}{r^2}-\frac{\log r}{2r^2}+\mathcal{O}\left(\frac{\log r}{r^3}\right),\\
V_4&\longrightarrow\frac{v_{_4}(\omega,q^2)}{r^2}+\mathcal{O}\left(\frac{1}{r^3}\right).
\end{split}
\end{equation}
Eqs.~(\ref{S1V2}, \ref{S2V1V3}, \ref{S3V4}) were used in order to obtain (\ref{boundary behavior}). The  boundary condition~(\ref{AdS constraint}) was also imposed.
From~(\ref{jmu formula2}), the current $J^{\mu}$ reads
\begin{equation}\label{jmu original}
\begin{split}
J^t=&\,\rho+2s_{_1}A_t^{(0)}+2s_{_2}\partial_k A_k^{(0)}+2s_{_3} \rho+ \frac{1}{2}q^2A_t^{(0)}-\frac{1}{2}i\omega \partial_k A_k^{(0)},\\
J^i=&\,-\left[2v_{_1}+\frac{1}{2}\left(\omega^2+q^2\right)\right]A_i^{(0)}-\left(2v_{_2}- \frac{1}{2}i\omega\right)\partial_iA_t^{(0)}\\
&\,-\left(2v_{_3}+\frac{1}{2}\right)\partial_i \partial_kA_k^{(0)}-2v_{_4}\partial_i \rho.
\end{split}
\end{equation}
The Landau frame convention~(\ref{landau frame}) requires (see Appendix~\ref{appLF} where  this condition is relaxed)
\begin{equation}\label{lf}
s_{_1}=-\frac{1}{4}q^2,~~~s_{_2}=\frac{1}{4}i\omega,~~~s_{_3}=0.
\end{equation}
The remaining coefficients ($v_{_i}$) cannot be fixed from pre-asymptotic considerations only and have to be found through integration of the dynamical equations~(\ref{S1V2},~\ref{S2V1V3},~\ref{S3V4}), from the horizon to the boundary.

We, however, were able to find two relations among the coefficients $S_i$ and $V_i$, which help to reduce the number of unknown coefficients.
Consider the combinations
\begin{equation}
X=i\omega S_1+q^2 S_2, ~~~~~~~Y=i\omega V_2+q^2V_3-V_1,
\end{equation}
which satisfy two coupled {\em homogeneous} ordinary differential equations
\begin{equation}
\left\{
\begin{aligned}
0=&r^2\partial_r^2X+3r\partial_rX-q^2\partial_rY,\\
0=&(r^5-r)\partial_r^2Y+(3r^4+1)\partial_rY-2i\omega r^3\partial_rY-i\omega r^2Y -r^3 \partial_rX-r^2X.
\end{aligned}
\right.
\end{equation}
$X$ and $Y$ have to vanish under our boundary conditions. Thus, two relations emerge
\begin{equation}\label{constraint relations1}
i\omega S_1+q^2 S_2=0, ~~~~~~~~~~i\omega V_2+q^2V_3-V_1=0
\end{equation}
At $r\rightarrow \infty$, the first relation is already imposed by the Landau frame choice (\ref{lf}).  The second one  is
\begin{equation}\label{constraint relations2}
i\omega v_{_2}+q^2 v_{_3}-v_{_1}=0,
\end{equation}
which will be used to eliminate $v_{_1}$ in favour of the others.


Finally, the current $J^i$ expressed  in a gauge invariant form reads
\begin{equation}
J^i=-2v_{_4}\partial_i \rho+\left(2v_{_2}-\frac{1}{2}i\omega\right)F_{ti}^{(0)}+ \left(2v_{_3}+ \frac{1}{2}\right) \partial_kF_{ki}^{(0)}.
\end{equation}
The transport coefficients $\mathcal{D}$, $\sigma_e$ and $\sigma_m$ are
\begin{equation}\label{Dv}
\mathcal{D}=2v_{_4},~~~\sigma_e=2v_{_2}-\frac{1}{2}i\omega,~~~\sigma_m=2v_{_3}+\frac{1}{2}.
\end{equation}
So far, we have only expressed the transport coefficients in terms of  the coefficients $v_{_i}$. The latter are yet to
be determined by solving the ODEs (\ref{S1V2}, \ref{S2V1V3}, \ref{S3V4}). This will be our goal in the next section.
We will first solve these equations analytically in the hydrodynamic limit $\omega\ll1,{q}\ll 1$ and then numerically for generic values of $\omega$ and $q$.

\section{Boundary current:  Results}\label{subsection42}

In this section, we  solve the equations~(\ref{S1V2},\ref{S2V1V3},\ref{S3V4}) and determine the transport coefficients, first perturbatively and then numerically.

\subsection{Perturbative analytic results: hydrodynamic expansion}\label{subsection51}

In the hydrodynamic limit $\omega\ll1,{q}\ll 1$, the equations~(\ref{S1V2},\ref{S2V1V3},\ref{S3V4}) could be solved perturbatively.
To this goal, we introduce an expansion parameter $\lambda$ via $\omega\to \lambda \omega$, $\vec{q}\to \lambda \vec{q}$.
Then,  $S_i$ and $V_i$ are formally expandable in series of $\lambda$,
\begin{equation}\label{lexp}
S_i(r,\omega,\vec{q})=\sum_{n=0}^{\infty}\lambda^nS_i^{(n)}(r,\omega,\vec{q}),~~~~ V_i(r,\omega,\vec{q})=\sum_{n=0}^{\infty}\lambda^nV_i^{(n)}(r,\omega,\vec{q}).
\end{equation}
$\lambda$  keeps track of the gradient expansion order and, eventually, will be set to unity.
The expansion (\ref{lexp}) is substituted into (\ref{S1V2},\ref{S2V1V3},\ref{S3V4}), and the equations are solved iteratively, order by order in $\lambda$.
At each order in $\lambda$,  direct integration results in the expansion coefficients $S_i^{(n)}$ and $V_i^{(n)}$ being expressed as certain double integrals.
All the details of this calculation are summarised in Appendix~\ref{appendix}. At $r\rightarrow\infty$,  the near boundary data $v_{_i}$ are extracted perturbatively
from $V_i^{(n)}$:
\begin{equation}
\begin{split}
v_{_1}&=\frac{1}{2}i\omega -\frac{1}{4}\left[q^2+\omega^2\left(1+\log 2\right)\right]+ \cdots,\\
v_{_2}&=\frac{1}{2}+\frac{1}{4}i\omega \left(1+\log 2\right)+\frac{1}{48}\left(\pi^2 \omega^2 -3\pi q^2-6q^2 \log 2\right)+\cdots,\\
v_{_3}&=-\frac{1}{4}+\frac{1}{32}i\omega \left(2\pi-\pi^2+4\log 2\right)+\cdots,\\
v_{_4}&=\frac{1}{4}+\frac{\pi}{16}i\omega -\frac{1}{96}\left[\pi^2\omega^2-q^2\left(6\log 2-3\pi\right)\right]+\cdots.
\end{split}
\end{equation}
When substituted into~(\ref{Dv}), these are the results quoted in~(\ref{hydro expansion}).

\subsection{Numerical results: all-order derivative resummation}\label{subsection52}
To all order in the gradient resummation, the equations~(\ref{S1V2},\ref{S2V1V3},\ref{S3V4}) have to be solved for generic values of $\omega$ and $q^2$, which  we have been able to do numerically only. We are facing  a boundary value problem for a system of second order ODEs. Much like what was done in~\cite{1406.7222,1409.3095,1502.08044,1504.01370}, we apply the shooting technique. We briefly sketch the numerical procedure. First, take any trial initial values for functions $V_i$ and $S_i$ at the horizon $r=1$ and integrate the equations up to the boundary $r=\infty$. The solution generated in this way has to satisfy the requirements~(\ref{AdS constraint}) and~(\ref{lf}) at the conformal boundary.  If it does not, the trial functions must be adjusted and the procedure is repeated until the boundary condition is fulfilled with a reasonably good numerical accuracy. This fine-tuning process of finding the correct initial  data is reduced to root-finding problem, which we solve by the Newton's method.

\subsubsection*{Diffusion}
\begin{figure}[htbp]
\centering
\includegraphics[scale=0.8]{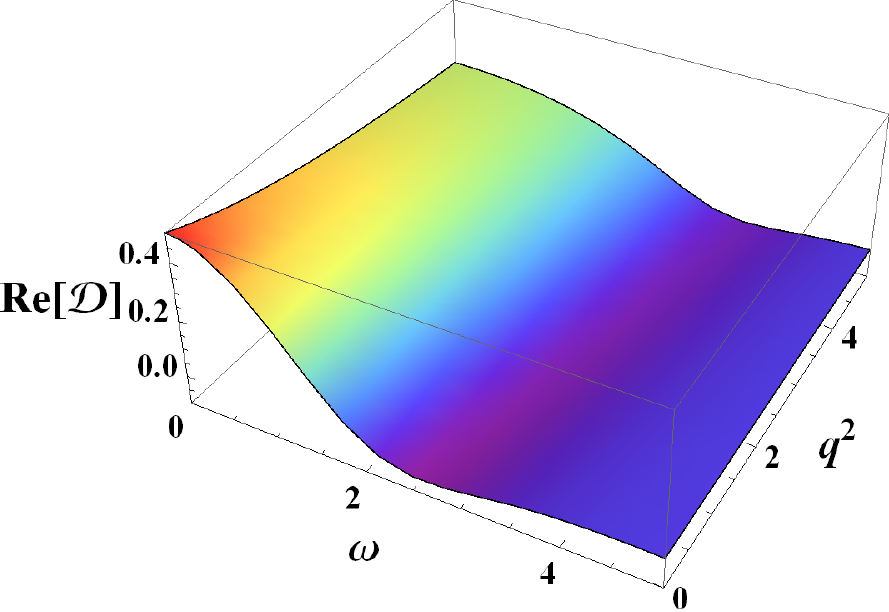}
\includegraphics[scale=0.8]{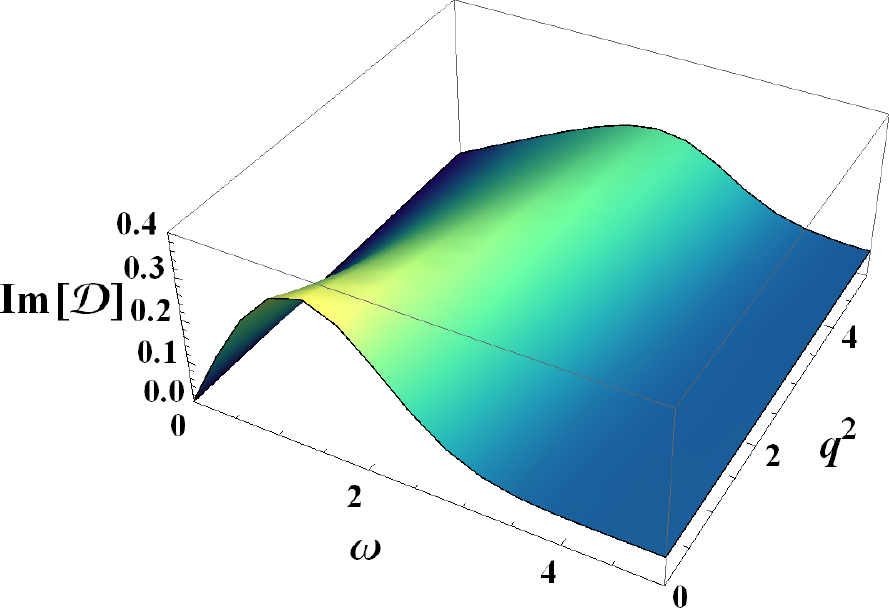}
\caption{Diffusion function $\mathcal{D}$ as a function of $\omega$ and $q^2$.}
\label{figure1}
\end{figure}

\begin{figure}[htbp]
\centering
\includegraphics[scale=0.8]{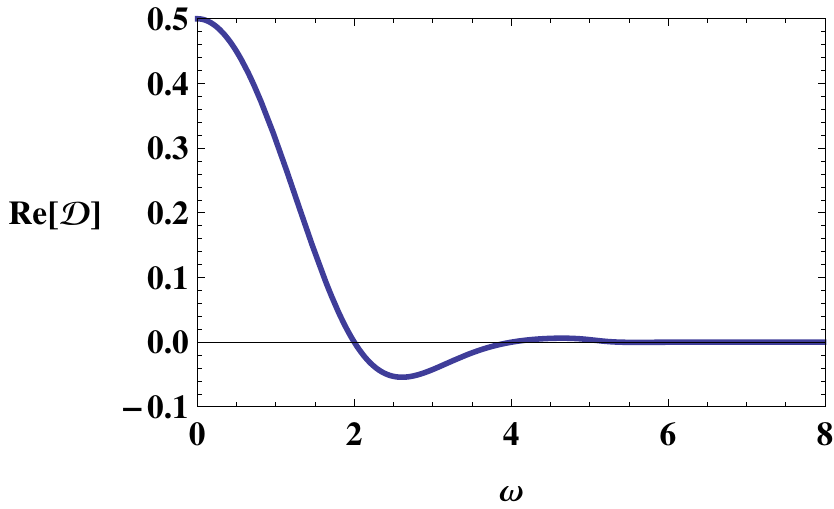}
\includegraphics[scale=0.8]{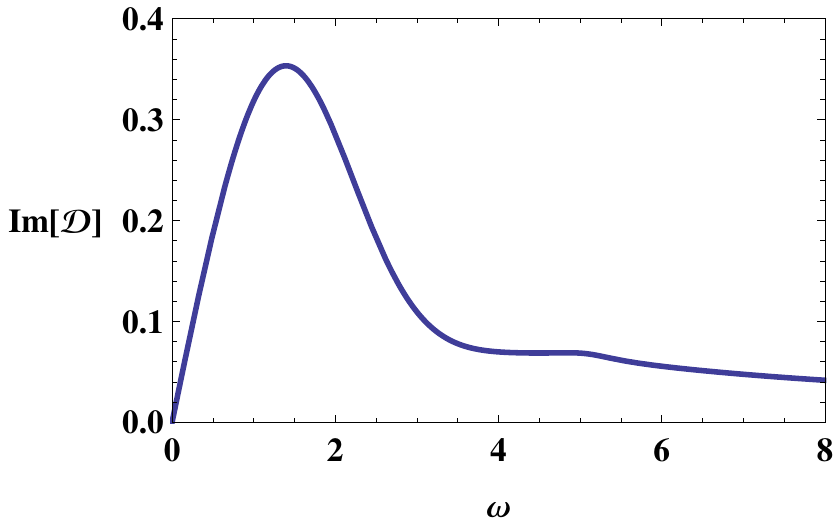}
\includegraphics[scale=0.8]{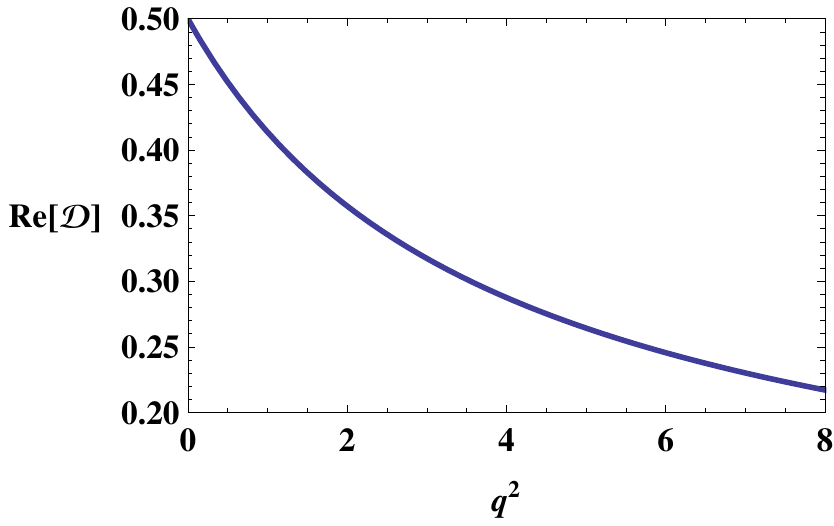}
\caption{Diffusion function $\mathcal{D}$ as a function of $\omega$ at $q=0$ (top); as a function of $q^2$ at $\omega=0$ (down).}
\label{figure3}
\end{figure}

We start presenting our numerical results with 3D plots for the diffusion function $\mathcal{D}$ (Figure~\ref{figure1}).
In figures~\ref{figure3}, we plot 2D sections for $\omega=0$ and $q=0$, which are more transparently revealing the asymptotic behaviour of $\mathcal{D}$.
Imaginary part of $\mathcal{D}$ vanishes at $\omega=0$.
As seen from the plots, $\mathcal{D}$ is a very smooth function of $q^2$. The weak dependence on $q^2$ reflects quasi-locality of the diffusion process.  The $\omega$
dependence is much more interesting.   The imaginary part of  $\mathcal{D}$ first approaches a maximum at $\omega\simeq1.5$, which could be interpreted as an emerging
scale in the problem. Both the real and imaginary parts display damped oscillations as a function of $\omega$.  This behaviour indicates  the presence of a set of complex poles, which are the quasi-normal modes of the bulk theory.
Importantly,  $\mathcal{D}$ eventually vanishes at large momenta. This behaviour is consistent with restoration of causality.

\begin{figure}[htbp]
\centering
\includegraphics[scale=0.8]{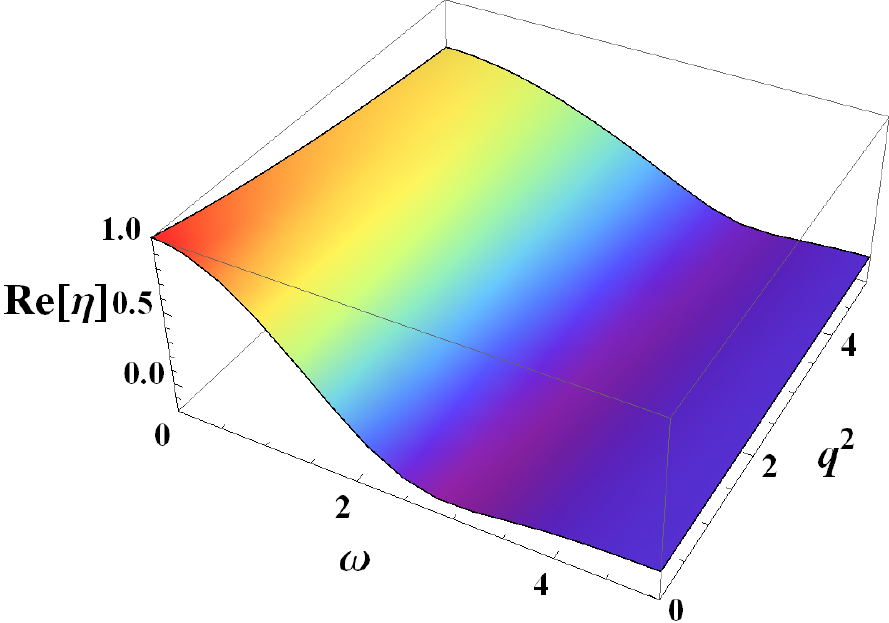}
\includegraphics[scale=0.8]{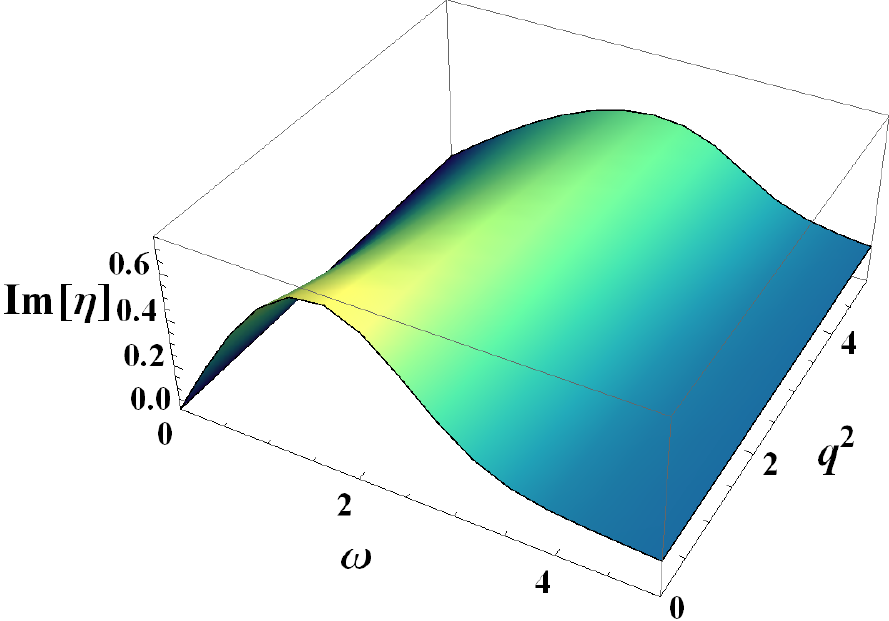}
\caption{Viscosity $\eta$ as a function of $\omega$ and $q^2$, taken from~\cite{1406.7222,1409.3095}.}
\label{figure2}
\end{figure}

It is interesting to notice that the diffusion function $\mathcal{D}$  behaves much like the viscosity $\eta$. The latter
was computed in~\cite{1406.7222,1409.3095}.  For the sake of comparison, we quote the results here  (figure~\ref{figure2}).
Roughly speaking, $\eta\sim 2\mathcal{D}$. It is not exact, however, as
is obvious from  the small momenta expansion for the viscosity $\eta$
\begin{equation}
\eta=1+\frac{1}{2}\left(2-\log 2\right)i\omega-\frac{1}{8}q^2-\frac{1}{48}\left[6\pi - \pi^2+12\left(2-3\log 2+\log^2 2\right)\right]\omega^2+\cdots\,,
\end{equation}
which is not proportional to one of $\mathcal{D}$.  In principle, there seems to be no reason why these two transport coefficient functions
should be proportional. Yet, the overall similarity in the behaviour is quite striking suggesting a sort of universality in dissipative transport.

\begin{figure}[htbp]
\centering
\includegraphics[scale=0.75]{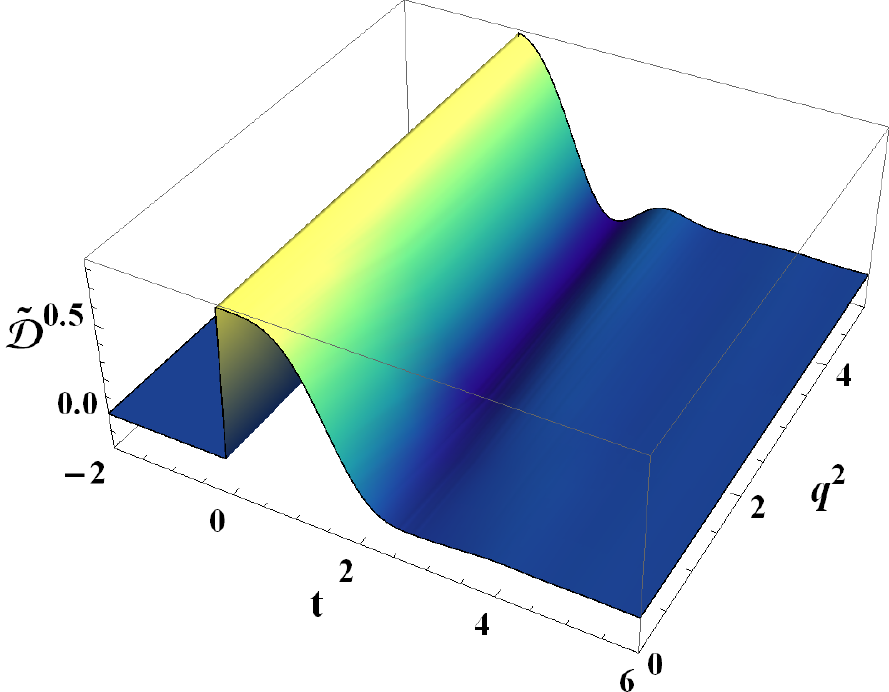}
\includegraphics[scale=0.8]{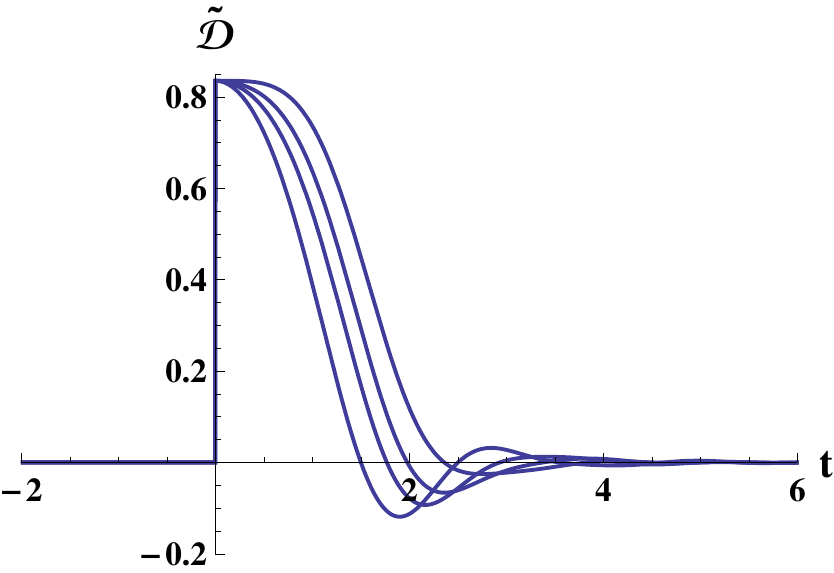}
\caption{Memory function $\widetilde{\mathcal{D}}(t,q^2)$. Left: 3D plot; Right: 2D slices of 3D plot with different curves corresponding to $q^2=0,1,2,4$ from the outmost.}
\label{figure4}
\end{figure}

Vanishing of $\mathcal{D}$ at large $\omega$ signals recovery of causality but does not imply it automatically.
To deeper explore the problem, we follow~\cite{1502.08044} and turn to the memory function formalism. The memory function $\widetilde{\mathcal{D}}$ is defined in~(\ref{memory fun1}).
It is displayed on figure~\ref{figure4}.
As anticipated, causal $\widetilde{\mathcal{D}}$ indeed does not have support at negative times (up to numerical noise).

\subsubsection*{Conductivity}
\begin{figure}[htbp]
\centering
\includegraphics[scale=0.8]{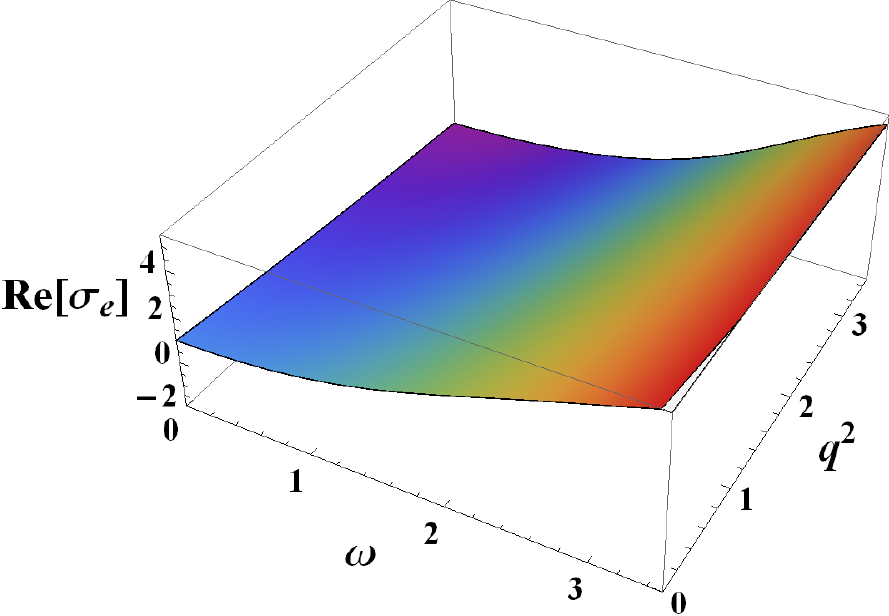}
\includegraphics[scale=0.8]{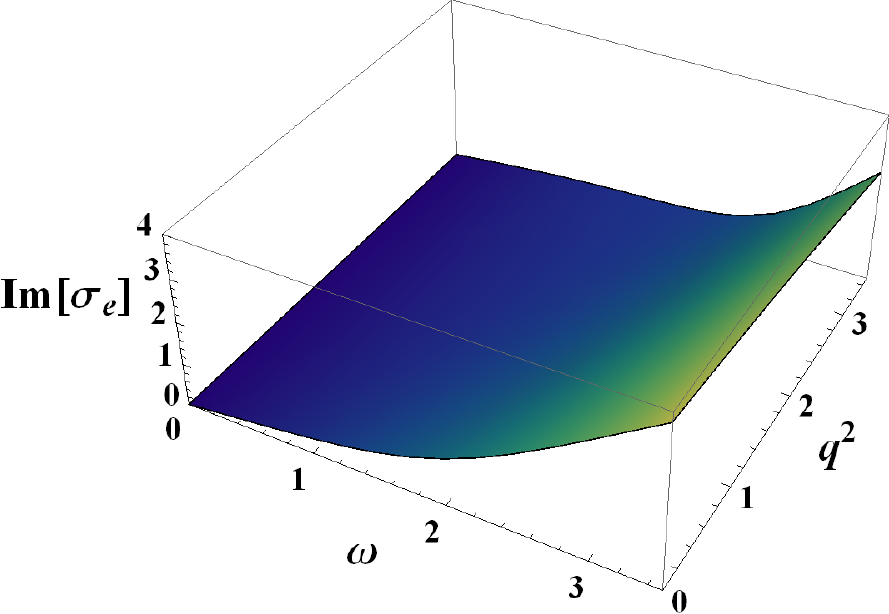}
\caption{Conductivity $\sigma_e$ as a function of $\omega$ and $q^2$.}
\label{figure5}
\end{figure}

\begin{figure}[htbp]
\centering
\includegraphics[scale=0.8]{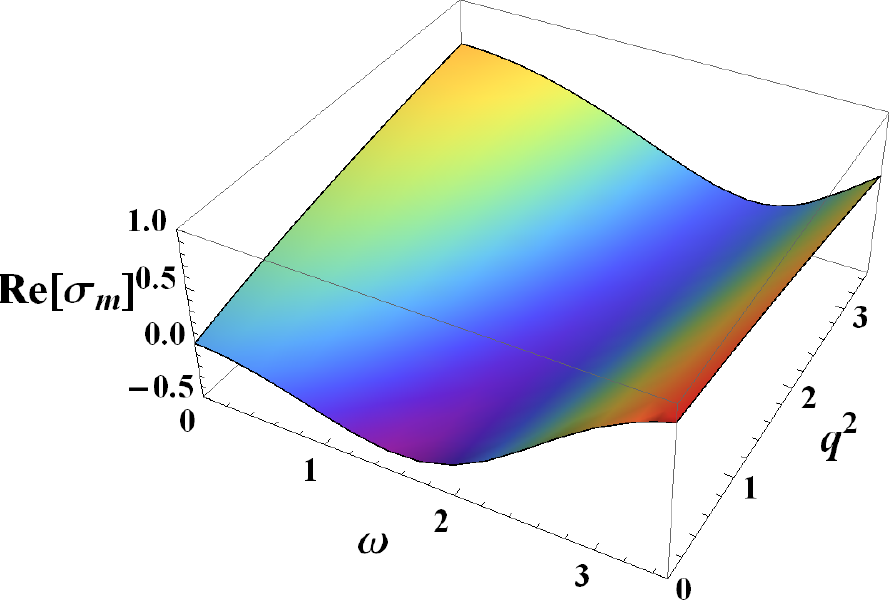}
\includegraphics[scale=0.8]{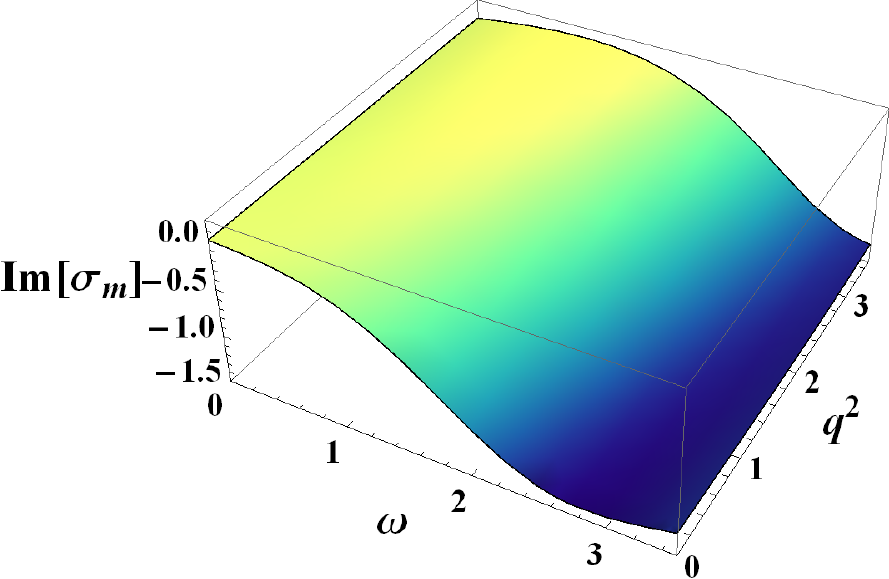}
\caption{Conductivity $\sigma_m$ as a function of $\omega$ and $q^2$.}
\label{figure6}
\end{figure}

\begin{figure}[htbp]
\centering
\includegraphics[scale=0.8]{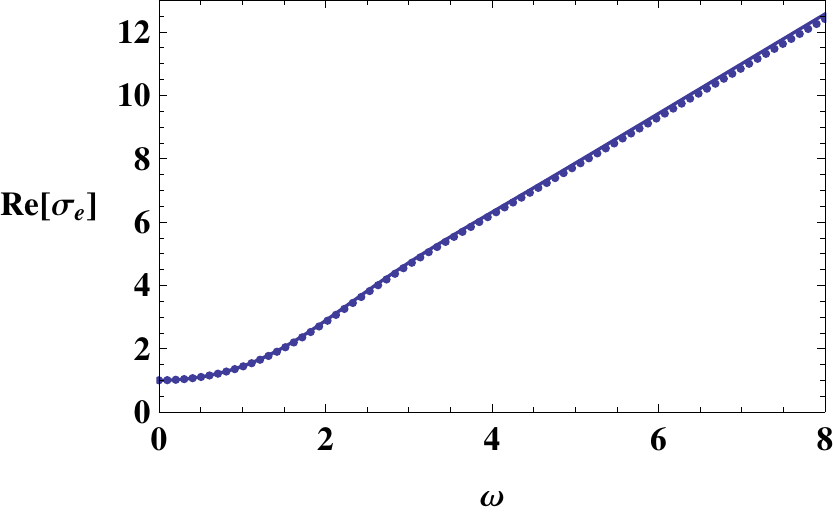}
\includegraphics[scale=0.8]{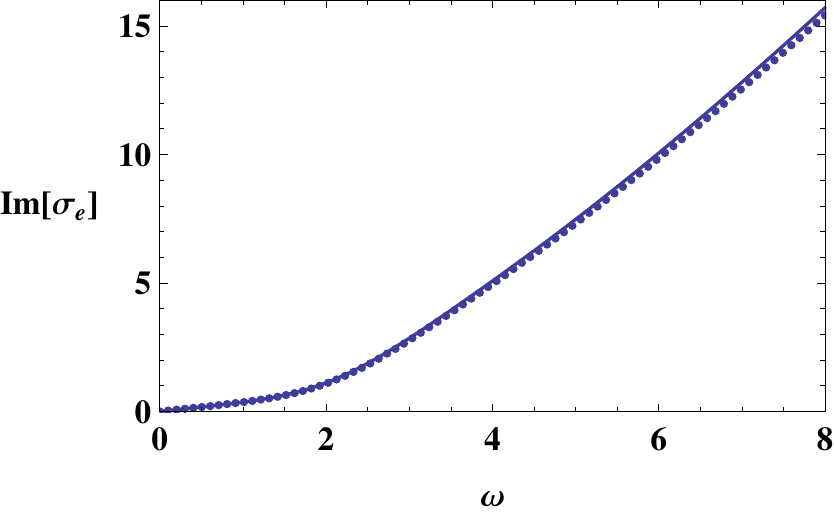}
\includegraphics[scale=0.8]{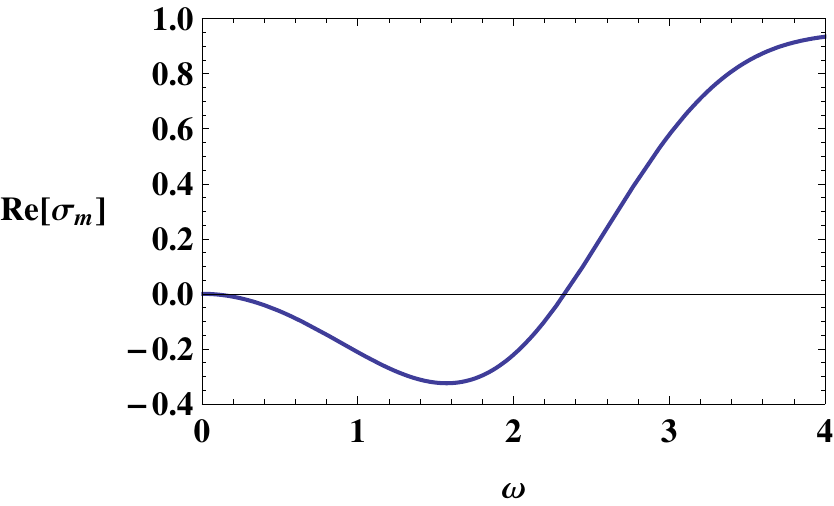}
\includegraphics[scale=0.8]{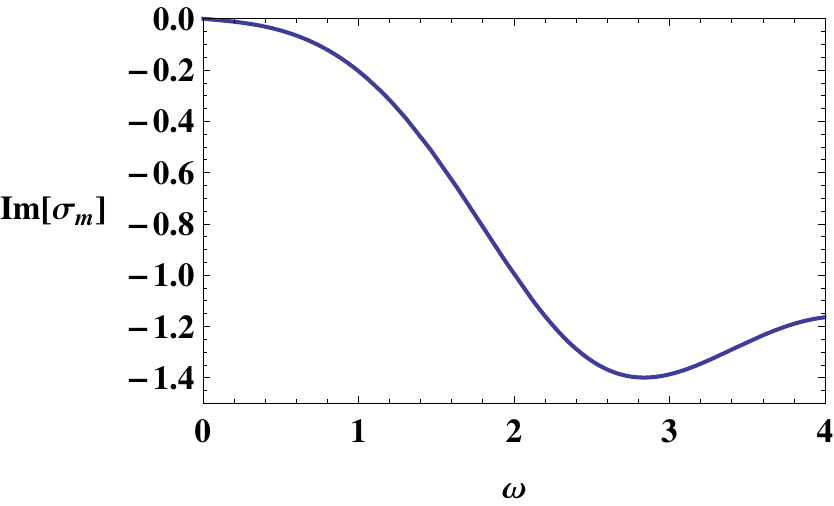}
\caption{Conductivities $\sigma_{e,m}$ as a function of $\omega$ at $q=0$. Top: The dashed line is our numerical results, while the solid line is the analytical result of~\cite{0810.1077}.}
\label{figure7}
\end{figure}

The results on the conductivities are shown in figures~\ref{figure5},~\ref{figure6} (3D), and~\ref{figure7} (2D). Similarly to the case of diffusion, the dependence on the spatial momentum $q$ is much less pronounced compared to the frequency $\omega$. The conductivity $\sigma_e$ is a growing function of $\omega$. It will become clear from the next subsection that at $q=0$, $\sigma_e$ coincides with the current-current two-point correlator computed analytically in~\cite{0810.1077,0706.0162},
\begin{equation}\label{analyt cond1}
\sigma_e(\omega,q=0)=i\omega \left\{\frac{1}{2}\psi \left[\frac{1}{4}(1-i)\omega \right] +\frac{1}{2}\psi\left[-\frac{1}{4}(1+i)\omega\right]+\frac{1}{2}\log 2+ \gamma_{_\textrm{E}}\right\}-1,
\end{equation}
where $\psi(x)$ is the digamma function $\psi(x)=\Gamma'(x)/\Gamma(x)$ and $\gamma_{_\textrm{E}}$ is Euler constant. At large $\omega$, it asymptotically approaches $\textrm{Re}[\sigma_e]\sim \omega$ and $\textrm{Im}[\sigma_e]\sim \omega \log \omega$. On figure~\ref{figure7}, both the numerical and  analytical results are shown  indicating a perfect agreement between the two.

\subsection{Continuity equation and two-point correlators}\label{subsection43}
In section~\ref{section4}, the dynamical components of~(\ref{Maxwell eq}) were solved for the correction $a_\mu$.
The resulting solutions for $a_\mu$ were expressed in terms of unspecified boundary fields $A_\nu^{(0)}$ and $\rho$. The fifth Maxwell equation is the constraint $\textrm{EQ}^r=0$
\begin{equation}\label{constraint eq}
\begin{split}
0=&r^2f(r)\partial_r\partial_ka_k-\partial^2 a_t+\partial_t\partial_ka_k-r^2 \partial_t \partial_r a_t -\partial^2A_t^{(0)}-\frac{1}{2r^2}\partial^2\rho+\partial_t\partial_k A_k^{(0)}+ \frac{1}{r}\partial_t\rho.
\end{split}
\end{equation}
Upon explicit use of the solution~(\ref{basis}) for $a_\mu$, it is easy to see that near the boundary $r\rightarrow\infty$ equation~(\ref{constraint eq}) reduces to the continuity equation $\partial_\mu J^\mu=0$.

\begin{figure}[htbp]
\centering
\includegraphics[scale=0.8]{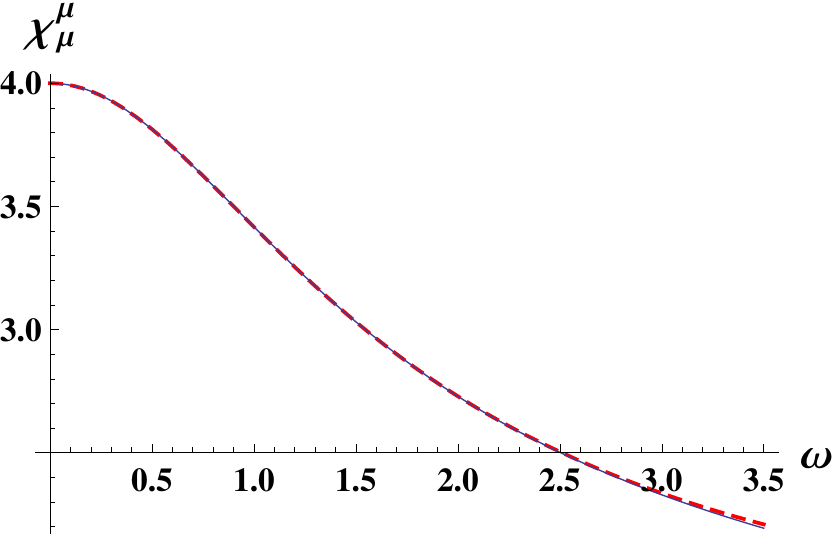}
\caption{Spectral function $\chi^{\mu}_{\mu}$ as a function of $\omega$ for light-like momenta $q=\omega$. The dashed line denotes our numerical results while the solid line is from~\cite{hep-th/0607237}.}
\label{figure12}
\end{figure}

The continuity equation makes it possible to eliminate $\rho$ in favour of the external electromagnetic potential $A_{\mu}^{(0)}$. The constitutive relation~(\ref{constitutive relation}) then takes the form of the usual linear response theory: $J^\mu=G^{\mu\nu}A_\nu^{(0)}$. In Fourier space, the current-current correlators $G^{\mu\nu}$ are
\begin{equation}\label{correlators}
\begin{split}
&G^{tt}=\frac{q^2\sigma_e}{i\omega-q^2\mathcal{D}},~~~~G^{it}=G^{ti}=\frac{\omega q_i \sigma_e}{i\omega-q^2\mathcal{D}},\\
&G^{ij}=\left(-i\omega \sigma_e-q^2\sigma_m\right)\left(\delta_{ij}-\frac{q_i q_j} {q^2} \right)+\frac{\omega^2\sigma_e}{i\omega-q^2\mathcal{D}}\cdot \frac{q_iq_j}{q^2}.
\end{split}
\end{equation}
The transverse component of the correlators $G^{\mu\nu}$ was analytically computed in~\cite{hep-th/0607237} for $q=\omega$. The analytically computed spectral function
$\chi_{\mu}^{\mu}\equiv -2\textrm{Im}[\eta_{\mu\nu}G^{\mu\nu}]$ is confronted against our numerical calculation in figure~\ref{figure12}, clearly demonstrating high
accuracy of the latter.
The longitudinal response function has infinitely many poles, solutions of the dispersion equation
\begin{equation}\label{dispersion}
i\omega-\mathcal{D}(\omega, q^2) q^2=0.
\end{equation}
In the bulk, these poles  correspond to  the quasi-normal modes. They are absent in the transverse part of the correlator.
In the hydrodynamic limit, substituting  the expansion~(\ref{hydro expansion})  and solving the dispersion relation~(\ref{dispersion}) perturbatively,
we reproduce the lowest diffusive pole
\begin{equation}
\omega=-\frac{i}{2}q^2-\frac{i}{8}q^4\log 2 +\cdots,
\end{equation}
which is in perfect agreement with~\cite{hep-th/0205052}. Perturbative expressions for the correlators~(\ref{correlators}) are also in agreement with~\cite{hep-th/0205052}.

\section{Remark on electromagnetic properties of the medium: dielectric tensor and refractive index} \label{subsection44}
So far we have treated the external fields as non-dynamical. As was mentioned earlier, we can gauge these fields turning them into fully dynamical. Coupled with the induced current, for which the constitutive relation was constructed, the theory at hand becomes a self-consistent electrodynamics of a conducting medium. Electromagnetic properties of this medium are related to two-point correlators and in principle could be deduced from the literature. For self-completeness of our study we summarise them below.

For medium with spatial dispersion, the more appropriate way to describe its response to external fields is the $(\vec{E},~\vec{D},~\vec{B})$ approach, where $\vec{H}$ is set equal to $\vec{B}$. The macroscopic Maxwell equations are
\begin{equation}
\vec \nabla \cdot \vec D=4\pi \rho^{ext}, \ \ \ \ \ \ \vec \nabla\cdot \vec B=0, \ \ \ \ \ \ \vec \nabla \times \vec E = - \dot{\vec B}, \ \ \ \ \
\vec \nabla\times \vec B= 4\pi \vec J^{ext} + \dot{\vec D},
\end{equation}
supplemented by a constitutive relation for the matter
\begin{equation}
\vec{D}=\hat \epsilon\, \vec{E},
\end{equation}
where $\hat \epsilon$ is a generalised dielectric tensor, taking into account both the dielectric and magnetic response. In Fourier space, $\hat \epsilon$ is a function of both $\omega$ and $\vec{q}$. A comparative discussion between the $(\vec{E},~\vec{D},~\vec{B})$ approach and the  formalism based on  four fields---$(\vec{E},~\vec{D},~\vec{B},~\vec{H})$ can be found in the textbooks~\cite{ag,mm}.

For an isotropic medium, $\hat\epsilon$ is decomposed into longitudinal and transverse components
\begin{equation}
\epsilon_{ij}(\omega,\vec q)=\epsilon^{\textrm{T}}(\omega,q)\left(\delta_{ij}- \frac{q_iq_j}{q^2}\right)+ \epsilon^{\textrm{L}}(\omega,q)\frac{q_i q_j}{q^2}.
\end{equation}
The constitutive relation~(\ref{constitutive relation}) is the relation for the induced current
\begin{equation}\label{crind}
\vec{J}_{ind}=-\mathcal{D}\vec{\nabla}\rho_{ind}+\sigma_e \vec{E}+\sigma_m\vec{\nabla} \times \vec{B}.
\end{equation}
Using the on-shell condition (continuity equation) and the Faraday law\footnote{The Faraday law is a Bianchi identity.} (the third Maxwell equation), the constitutive relation~(\ref{crind}) becomes the Ohm's law for the induced current
\begin{equation}
\vec J_{ind}\,=\,\hat \sigma\,\vec E\,, \ \ \ \ \ \ \ \ \
\hat \sigma_{ij}(\omega,\vec q)=\sigma^{\textrm{T}}(\omega,q^2)\left(\delta_{ij}- \frac{q_iq_j}{q^2}\right)+ \sigma^{\textrm{L}}(\omega,q^2)\frac{q_i q_j}{q^2}.
\end{equation}
The longitudinal and transverse components of the conductivity are proportional to the current-current correlation function $G^{\mu\nu}$ and are trivially related to the transport coefficient functions of~(\ref{crind}):
\begin{equation}\label{cond TL}
\sigma^{\textrm{L}}(\omega,q)={i\omega \sigma_e(\omega,q) \over i\omega-q^2 \mathcal{D}(\omega,q) }\,, \ \ \ \ \ \ \ \ \ \
\sigma^{\textrm{T}}(\omega,q)= \sigma_e(\omega,q) - i{q^2\over \omega} \sigma_m(\omega,q).
\end{equation}
The dielectric functions are related to the conductivity via the relation:
\begin{equation}
\epsilon^{\textrm{L,T}}=1+\frac{4\pi i \sigma^{\textrm{L,T}}}{\omega}.
\end{equation}

\begin{figure}[htbp]
\centering
\includegraphics[scale=0.8]{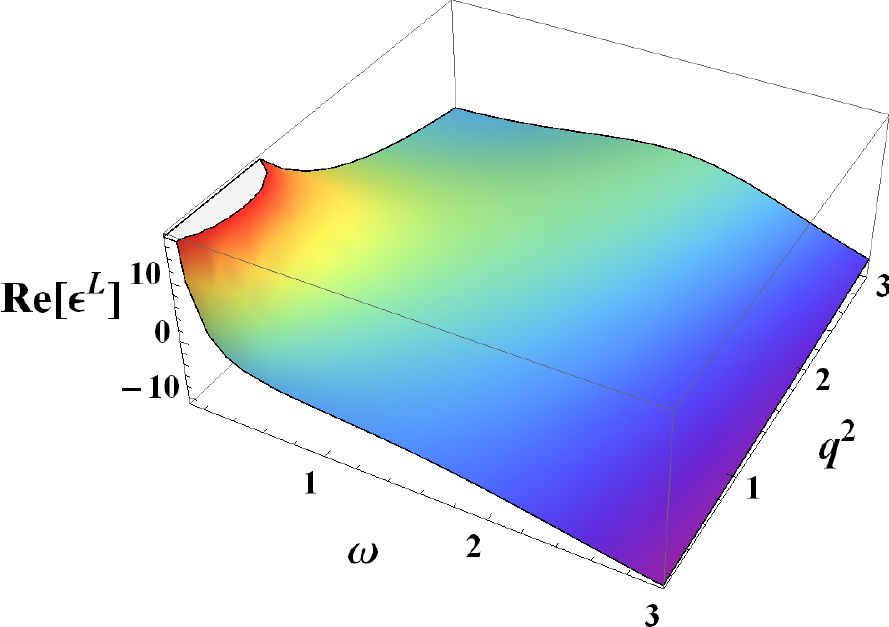}
\includegraphics[scale=0.8]{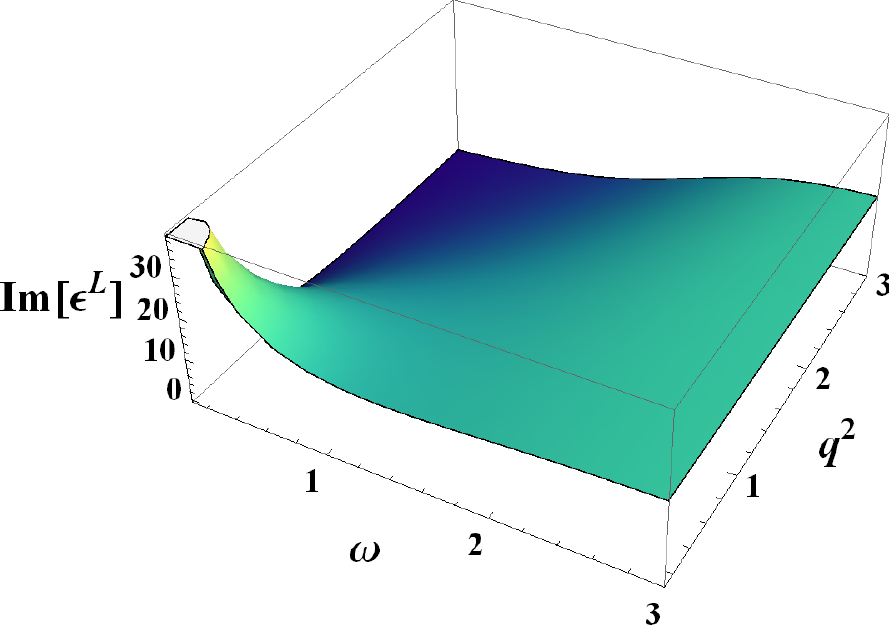}
\includegraphics[scale=0.8]{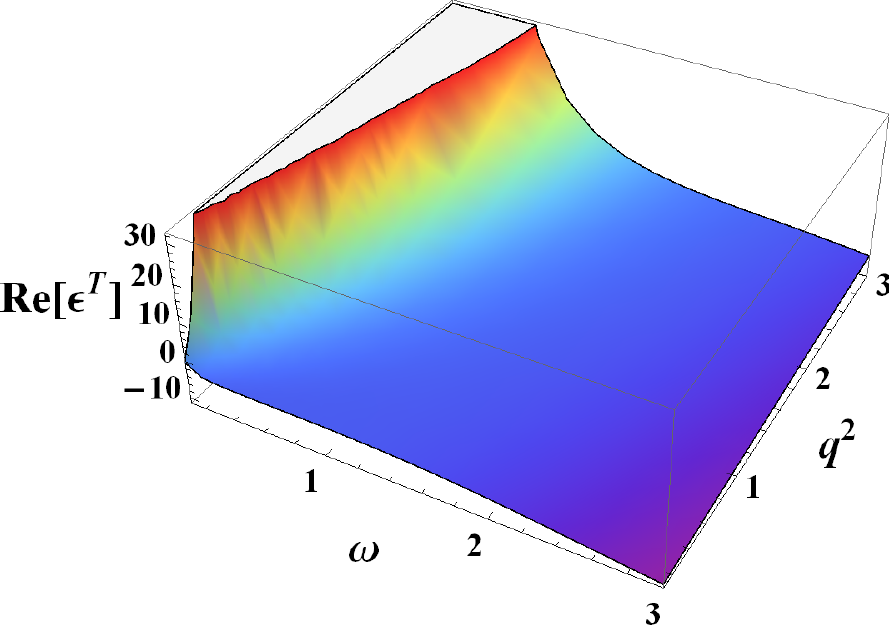}
\includegraphics[scale=0.8]{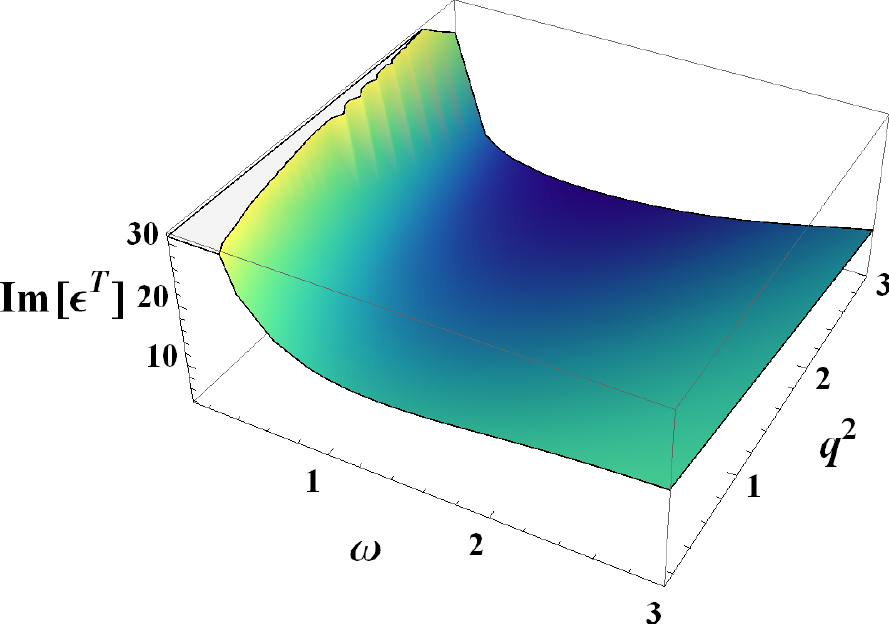}
\caption{Dielectric functions $\epsilon^{\textrm{L,T}}$ as a function of $\omega$ and $q^2$.}
\label{figure8}
\end{figure}

Numerical results for $\epsilon^{\textrm{L,T}}$ are shown in figure~\ref{figure8}. The electromagnetic waves propagating in the medium should satisfy dispersion relations
\begin{equation}\label{prop waves}
\begin{split}
&\textrm{Transvese mode}:~~~~~~~\frac{q^2}{\omega^2}=\epsilon^{\textrm{T}},\\
&\textrm{Longitudinal mode}:~~~\epsilon^{\textrm{L}}=0.
\end{split}
\end{equation}
The transverse mode defines the refractive index $n^2=\epsilon^{\textrm{T}}$, which has two branches for $n$. Depending on the sign of $\textrm{Im}[n]$, only one branch is physical.

The first three propagating waves for both transverse and longitudinal modes were computed numerically in~\cite{1404.4048}, revealing a rich structure of the electromagnetic response of such holographic medium. Remarkably, the authors of~\cite{1404.4048} found that one of the transverse modes displays negative refractive  phenomenon. Previous studies~\cite{1006.5714,1107.1240} noticed this phenomenon in the hydrodynamic regime of charged fluids which holographically are described by $R$-charge black holes. The criterion proposed in ref.~\cite{physics/0311029} for negative refraction was used in~\cite{1006.5714,1107.1240}. The latter criterion is, however, valid for perturbatively small $q$ only. The negative refraction mode of~\cite{1404.4048}  is gapped and cannot be treated perturbatively. This is the reason the authors of~\cite{1404.4048} relied on the condition $\textrm{Re}[n]<0$ ($\textrm{Im}[n]>0$) to establish the phenomenon.

We would like to remark that requiring negativity of $\textrm{Re}[n]$ might not be sufficient to make a conclusive statement about the existence of propagating modes with negative refraction. One must carefully study the energy propagation in the medium and establish direction of the Poynting vector.
The time-averaged Poynting vector is~\cite{ag}
\begin{equation}
\vec{P}=\frac{1}{2}\textrm{Re}\left(\vec{E}^*\times \vec{B}\right)- \textrm{Re}\left\{ \frac{\omega}{4} \frac{\partial\epsilon_{ij}}{\partial\vec{q}}\,E_i^* E_j\right\},
\end{equation}
where the second term is due to spatial dispersion and plays an important role for the existence of negative refraction phenomenon. For a specific mode $\omega_{_{\textrm{M}}}$ solving the dispersion relation~(\ref{prop waves}), the direction of $\vec P$ is defined by the sign of $\textrm{Re}\left(\omega_{_M}\,\frac{d\epsilon^T}{d\omega}\frac{d\omega_{_M}}{dq}\right)$. In general, $\omega_{_M}$ is a complex valued function of $q$ turning this study into quite complicated. If its imaginary part can be neglected (transparent medium), one can show that $\vec{P}$ is directed along the group velocity~\cite{ag}
\begin{equation}
\vec{P}=U\vec{v}_{\textrm{gr}},~~~\vec{v}_{\textrm{gr}}=\frac{d\omega_{_{\textrm{M}}}(q)}{dq} \frac{\vec{q}}{q}
\end{equation}
where
$U$ is the electromagnetic wave energy density. While refs.~\cite{1006.5714,1107.1240,1404.4048} have made a significant progress towards establishing the negative refraction in holographic fluids, we feel that a further exploration of this phenomenon is worth pursuing.
\section{Summary}\label{summary}

The most general off-shell constitutive relation for a globally conserved $U(1)$ current~(\ref{constitutive relation}) can be parametrised using
three independent momenta-dependent transport coefficient functions: the diffusion $\mathcal{D}(\omega,q^2)$ and two conductivities $\sigma_{e,m}(\omega,q^2)$. The latter couple the current to external electromagnetic fields. At small momenta, these functions are power expandable reflecting the gradient expansion of the hydrodynamic regime.
In contrast, causality of the theory is manifested in the large momentum asymptotics.

Precise determination of all three coefficient functions using the fluid-gravity correspondence is our main result. In the hydrodynamic regime, we analytically reproduced all known results on the gradient expansion and extended them to the third order~(\ref{hydro expansion}). Particularly, for the first time an analytic expression was given for $\sigma_m$. The rest of the results were obtained numerically, beyond hydrodynamic regime. Of particular interest is the large $\omega$ behaviour, which for the first time was shown to be consistent with the causality constraints. We also studied causality using a numerically computed memory function. The memory function $\widetilde{\mathcal{D}}(t)$ was demonstrated to have no support for negative times thus further revealing causality restoration due to all order derivatives resummation.

We have noticed  an interesting similarity between  the diffusion $\mathcal{D}(\omega,q^2)$ and the viscosity $\eta(\omega,q^2)$: both functions demonstrate a strikingly similar behaviour as functions of $\omega$ and $q$ (even though  they are not exactly equal). We speculate that this might point to a certain universal
behaviour among various dissipative transport coefficient functions.

The transport coefficients were deduced using the holographic prescription. In the dual gravity side, we constructed a general solution for a probe Maxwell field in  Schwarzschild-$AdS_5$ black hole spacetime.
The Maxwell equations in the bulk were split into dynamical and constraint components. Interestingly, the off-shell constitutive relation~(\ref{constitutive relation}) and the corresponding transport coefficient functions were found to be fixed uniquely through solutions of the dynamical components only.
The constraint component was shown to be equivalent to the continuity equation $\partial_\mu J^\mu=0$.

As yet another check of our formalism, we used the continuity equation (on-shell condition) to reproduce some known results on two-point correlators~(\ref{correlators}).
It is, however, worth emphasising that in linear response there are only two independent two-point correlators (transverse and longitudinal), whereas our formalism reveals three fully independent transport coefficients. Our off-shell structure emerges as an important element of hydrodynamic effective action~\cite{1511.03646} and is clearly more rich than what is encoded in the on-shell correlators.



\appendix

\section{Relaxing the Landau frame condition}\label{appLF}

In this Appendix, we demonstrate that  all the transport coefficients can be uniquely fixed without imposing the
Landau frame convention~(\ref{landau frame}). The latter  is equivalent to fixing
the residual gauge, whereas the physical quantities (transport coefficients) are gauge invariant.

The gauge invariance of $J^{\mu}$ (eq.~(\ref{jmu original})) under boundary gauge transformation
\begin{equation}\label{bdry gauge}
J^\mu\to J^\mu~~~\textrm{under}~~~A_\mu^{(0)}(x_\alpha)\to A_\mu^{(0)}(x_\alpha) +\partial_\mu\varphi(x_\alpha),
\end{equation}
 gives the  relations
\begin{equation}\label{cons relation}
i\omega s_{_1}+q^2 s_{_2}=0,~~~~i\omega v_{_1}+q^2v_{_3}-v_{_1}=0.
\end{equation}
Along with the condition~(\ref{AdS constraint1}) and regularity requirement for $S_i,V_i$ at the horizon, the relations~(\ref{cons relation})
are equivalent  to the constraints~(\ref{constraint relations1}). The constitutive relation for $J^i$ of~(\ref{jmu original}) in a gauge invariant form is
\begin{equation}
\begin{split}
J^i=&\;-\frac{2v_{_4}}{1+2s_{_3}}\partial_iJ^t+\left[\left(2v_{_2}- \frac{1}{2} i\omega \right)-\frac{2v_{_4}} {1+2s_{_3}} \left(2 s_{_1}+ \frac{1}{2}q^2\right) \right] \frac{ \partial_i \partial_kE_k}{\partial^2}\\
&+\frac{2v_{_1}+(\omega^2+q^2)/2}{i\omega}\left(E_i-\frac{\partial_i\partial_kE_k} {\partial^2}\right)
\end{split}
\end{equation}
Here $\rho$ was exchanged in favour of $J^t$ and the Bianchi identity $\vec{\nabla}\times \vec{E}=-\partial_t \vec{B}$
was used to substitute $E_i$ for $B_i$. The transport coefficients of interest are
\begin{equation}
\begin{split}
&\mathcal{D}=\frac{2v_{_4}}{1+2s_{_3}},~~~~\bar{\sigma}^{\textrm{T}}=\frac{2v_{_1}+ (\omega^2+q^2)/2}{i\omega},\\
&\bar{\sigma}^{\textrm{L}}=\left(2v_{_2}- \frac{1}{2} i\omega \right)-\frac{2v_{_4}} {1+2s_{_3}} \left(2 s_{_1}+ \frac{1}{2}q^2\right).
\end{split}
\end{equation}
The coefficients $\sigma_e$ and $\sigma_m$ are linear combinations of $\bar{\sigma}^{\textrm{T}}$ and $\bar{\sigma}^{\textrm{L}}$.
The equation for $V_1$ is decoupled from all the other equations. Moreover,
the asymptotic boundary condition~(\ref{AdS constraint1}) and regularity at horizon are sufficient to fully determine $V_1$.  Thus its pre-asymptotic value $v_{_1}$
and hence the transport coefficient $\bar{\sigma}^{\textrm{T}}$ are uniquely fixed and just the ones quoted in the text.

The situation with the coefficients $\mathcal{D}$ and $\bar{\sigma}^{\textrm{L}}$ is more involved.
Below we will prove that $\mathcal{D}$ and $\bar{\sigma}^{\textrm{L}}$ are also uniquely fixed even without the Landau frame condition~(\ref{lf}) imposed.

First, consider $\mathcal{D}=2v_{_4}/(1+2s_{_3})$ which is to be determined from the equations~(\ref{S3V4}). Through the transformation
\begin{equation}
\tilde{S}_3=S_3+\frac{1}{2r^2},
\end{equation}
the equations~(\ref{S3V4}) turn into homogeneous
\begin{equation}\label{S3V4tr}
\left\{
\begin{aligned}
0=&r^2\partial_r^2\tilde{S}_3+3r\partial_r\tilde{S}_3-q^2\partial_rV_4,\\
0=&(r^5-r)\partial_r^2V_4+(3r^4+1)\partial_rV_4-2i\omega r^3\partial_rV_4-i\omega r^2V_4 -r^3\partial_r\tilde{S}_3-r^2\tilde{S}_3.
\end{aligned}
\right.
\end{equation}
Near $r=\infty$, the boundary conditions are not modified:
\begin{equation}\label{bdry S3V4tr}
\tilde{S}_3\to 0,~~~~V_4\to 0.
\end{equation}
Regularity near horizon  constrains $V_4$ only. So, the solutions to~(\ref{S3V4tr}) are parameterised by a  choice of $s_{_3}$. Homogeneity property of~(\ref{S3V4tr}), along with the boundary conditions~(\ref{bdry S3V4tr}) admit a family of solutions which have $r$-independent scaling symmetry. That is, only the ratio $V_4/\tilde S_3$ is fixed.  At the boundary this translates into $\mathcal{D}=2v_{_4}/(1+2s_{_3})$ being independent of the $s_{_3}$ choice and is thus uniquely fixed. Fixing $s_{_3}$, particularly taking $s_{_3}=0$ is just a convenient choice utilizing this freedom.

Next, we turn to $\left\{S_1,V_2\right\}$ from the equations~(\ref{S1V2}). Through the shift
\begin{equation}
\tilde{S}_1\to S_1+1,
\end{equation}
we have homogeneous equations for $\tilde{S}_1$ and $V_2$
\begin{equation}\label{S1V2tr}
\left\{
\begin{aligned}
0=&r^2\partial_r^2\tilde{S}_1+3r\partial_r\tilde{S}_1-q^2\partial_rV_2,\\
0=&(r^5-r)\partial_r^2V_2+(3r^4+1)\partial_rV_2-2i\omega r^3\partial_rV_2-i\omega r^2V_2 -r^3\partial_r\tilde{S}_1-r^2\tilde{S}_1,
\end{aligned}
\right.
\end{equation}
which bear the same form as those of $\left\{\tilde{S}_3,V_4\right\}$. However, the boundary conditions near $r=\infty$ are now changed,
\begin{equation}\label{bdry S1V2tr}
\tilde{S}_1\to 1,~~~~~V_2\to 0.
\end{equation}
Regularity near horizon  effectively constrains $V_2$ only.  Solutions to~(\ref{S1V2tr}) are parameterised by  $s_{_1}$. However, contrary to the case of $\tilde S_3,V_4$, we do not have a freedom to rescale $\tilde S_1$ as this would modify the boundary conditions~(\ref{bdry S1V2tr}).

We proceed in the following way. Let split $\tilde{S}_1$ and $V_2$ into two pieces
\begin{equation}
\tilde{S}_1=\tilde{S}_1^{\Delta}+\Delta \tilde{S}_1,~~~~V_2=V_2^{\Delta}+\Delta V_2,
\end{equation}
where $\left\{\tilde{S}_1^{\Delta},V_2^{\Delta}\right\}$, $\left\{\Delta \tilde{S}_1,\Delta V_2\right\}$ obey the same equations as those of $\left\{\tilde{S}_1, V_2\right\}$. Near $r=\infty$, we specify the boundary conditions
\begin{equation}
\begin{split}
&\tilde{S}_1^{\Delta} \to 1,~~~~V_2^{\Delta} \to 0,\\
&\Delta \tilde{S}_1 \to 0,~~~~\Delta V_2\to 0,
\end{split}
\end{equation}
while at the horizon we impose regularity for all.
 Near $r=\infty$, we can have similar boundary expansions for $\left\{\tilde{S}_1^{\Delta }, V_2^{\Delta}\right\}$ and $\left\{\Delta \tilde{S}_1,\Delta V_2\right\}$, as those for $S_i,V_i$,
\begin{equation}
\begin{split}
\tilde{S}_1^{\Delta}&\longrightarrow1+\frac{\tilde{s}_{_1}^{\Delta }}{r^2}+\cdots,\\
V_2^{\Delta}&\longrightarrow -\frac{i\omega}{r}+\frac{v_{_2}^{\Delta }}{r^2}+\cdots,\\
\Delta \tilde{S}_1&\longrightarrow\frac{\Delta \tilde{s}_{_1}}{r^2}+\cdots,\\
\Delta V_2&\longrightarrow \frac{\Delta v_{_2}}{r^2}+\cdots.\\
\end{split}
\end{equation}

We can think of $\left\{\tilde{S}_1^{\Delta},V_2^{\Delta}\right\}$ as being uniquely fixed by some choice of $\tilde{s}_{_1}^{\Delta }$. The entire
uncertainty in  the solution is then shifted into the ``corrections'' $\left\{\Delta \tilde{S}_1,\Delta V_2\right\}$. Notice that these corrections have the same scaling symmetry as those of $\left\{\tilde{S}_3,V_4\right\}$: $\Delta V_2/{ \Delta \tilde{S}_1}=V_4/{ \tilde{S}_3}=\mathcal{D}$. This in fact proves that the combination $V_2 - \mathcal{D} S_1$ is uniquely fixed, without specifying $s_1$. Therefore, $\bar{\sigma}^{\textrm{L}}$ is independent of $s_{_1}$ and $s_{_3}$. This completes the proof.


\section{Perturbative solutions for $S_i$ and $V_i$}\label{appendix}

The coefficient functions  $S_i$ and $V_i$ can be formally expanded as
\begin{equation}
S_i=\sum_{n=0}^{\infty}\lambda^n S_i^{(n)},~~~~~V_i=\sum_{n=0}^{\infty} \lambda^n V_i^{(n)}.
\end{equation}
Then the eqs.~(\ref{S1V2},~\ref{S2V1V3},~\ref{S3V4}) are solved perturbatively in $\lambda$.
The results are summarised below.
\paragraph{I: $\left\{S_1,~V_2\right\}$}
\begin{equation}
S_1^{(0)}=S_1^{(1)}=0,
\end{equation}
\begin{equation}
\begin{split}
&V_2^{(0)}=-\frac{1}{4}\left(\pi-2 \arctan (r)+\log\frac{(1+r)^2}{1+r^2}\right) \longrightarrow -\frac{1}{r}+\frac{1}{2r^2}+\mathcal{O}\left(\frac{1}{r^3}\right),
\end{split}
\end{equation}
\begin{equation}
\begin{split}
&V_2^{(1)}=-\int_r^{\infty}\frac{xdx}{x^4-1}\int_1^xdy\left[2i\omega y\partial_y V_2^{(0)}(y)+i\omega V_2^{(0)}(y)\right]\\
&~~~~~~~~~\longrightarrow \frac{1}{4r^2}i\omega \left(1+\log 2-2\log r\right)+ \mathcal{O}\left(\frac{1}{r^3}\right),
\end{split}
\end{equation}
\begin{equation}
\begin{split}
&S_1^{(2)}=-\int_r^{\infty}\frac{dx}{x^3}\int_1^x q^2y\partial_y V_2^{(0)}(y) dy- \frac{1}{16r^2}q^2\left(\pi+6\log 2\right)\\
&~~~~~~~~~\longrightarrow -\frac{q^2}{4r^2}-\frac{q^2\log r}{2r^2}+ \mathcal{O} \left(\frac{1}{r^3}\right), \\
\end{split}
\end{equation}
\begin{equation}
\begin{split}
&V_2^{(2)}=-\int_r^\infty\frac{xdx}{x^4-1}\int_1^xdy\left[2i\omega y \partial_y V_2^{(1)}(y)+i\omega V_2^{(1)}(y)+y\partial_y S_1^{(2)}(y)+S_1^{(2)}(y)\right]\\
&~~~~~~~~~\longrightarrow \frac{1}{48r^2}\left[\pi^2\omega^2-3q^2\left(\pi+2\log 2\right)\right]+\mathcal{O}\left(\frac{1}{r^3}\right).
\end{split}
\end{equation}
\paragraph{II: $\left\{S_2,~V_1,~V_3\right\}$}
\begin{equation}
S_2^{(0)}=0,~~~~V_1^{(0)}=0,
\end{equation}
\begin{equation}
\begin{split}
&V_3^{(0)}=\frac{1}{16}\left\{-\pi^2-4\log r \left[-i\pi +\log\frac{r^2+1}{r^2-1} \right] +4\textrm{Li}_2(r^2)-\textrm{Li}_2(r^4)\right\}\\
&~~~~~~~~~\longrightarrow -\frac{1}{4r^2}-\frac{\log r}{2r^2}+\mathcal{O} \left(\frac{1}{r^3}\right),
\end{split}
\end{equation}
\begin{equation}
V_1^{(1)}=-\frac{1}{4}i\omega \left[\pi-2 \arctan (r)+\log\frac{(1+r)^2}{1+r^2}\right] \longrightarrow -\frac{i\omega }{r}+\frac{i\omega}{2r^2}+\mathcal{O} \left(\frac{1}{r^3}\right),
\end{equation}
\begin{equation}
\begin{split}
&S_2^{(1)}=-\int_r^\infty \frac{dx}{x^3}\int_1^x\left[-y\partial_y V_1^{(0)}(y)\right]dy- \frac{1}{16r^2}i\omega \left(-8+\pi +6\log 2\right)\\
&~~~~~~~~~\longrightarrow \frac{i\omega}{4r^2}-\frac{i\omega \log r}{2r^2}+\mathcal{O} \left(\frac{1}{r^3}\right),
\end{split}
\end{equation}
\begin{equation}
\begin{split}
&V_3^{(1)}=-\int_r^\infty \frac{x dx}{x^4-1}\int_1^x dy\left[2i\omega \partial_y V_3^{(0)}(y)+i\omega V_3^{(0)}(y)+y\partial_y S_2^{(1)}(y)+S_2^{(1)}(y)\right.\\
&~~~~~~~~\left.+\frac{V_1^{(1)}(y)}{y} \right] \longrightarrow \frac{i\omega}{32r^2}\left(2\pi-\pi^2+4\log 2\right)+\mathcal{O} \left(\frac{1}{r^3} \right),
\end{split}
\end{equation}
\begin{equation}
\begin{split}
&V_1^{(2)}=-\int_r^\infty \frac{x dx}{x^4-1}\int_1^r dy\left[2i\omega y \partial_y V_1^{(1)}(y)+i\omega V_1^{(1)}(y)+q^2y^{-1}\right]\\
&~~~~~~~~~\longrightarrow -\frac{1}{4r^2}\left[q^2+\omega^2\left(1+\log 2\right)\right]- \frac{1}{2r^2}\left(q^2-\omega^2\right)\log r +\mathcal{O}\left(\frac{1}{r^3}\right),
\end{split}
\end{equation}
\begin{equation}
\begin{split}
&V_1^{(3)}=-\int_r^\infty \frac{x dx}{x^4-1}\int_1^r dy \left[2i\omega y \partial_y V_1^{(2)}(y) +i\omega V_1^{(2)}(y)+\frac{q^2}{y}V_1^{(1)}(y)\right]\\
&~~~~~~~~~\longrightarrow \frac{1}{96r^2}\pi^2 i\omega \left(-3q^2+2\omega^2\right) +\mathcal{O}\left(\frac{1}{r^3}\right),
\end{split}
\end{equation}
where $\textrm{Li}_n(x)$ is the polylogarithm function.
\paragraph{III: $\left\{S_3,~V_4\right\}$}
\begin{equation}
S_3^{(0)}=S_3^{(1)}=0,
\end{equation}
\begin{equation}
V_4^{(0)}=\frac{1}{8}\left(\pi-2 \arctan(r)-\log\frac{(1+r)^2}{1+r^2}\right) \longrightarrow \frac{1}{4r^2}+\mathcal{O} \left(\frac{1}{r^3}\right),
\end{equation}
\begin{equation}
V_4^{(1)}=-\int_r^\infty\frac{x dx}{x^4-1}\int_1^rdy\left[2i\omega y \partial_y V_4^{(0)}(y)+i\omega V_4^{(0)}(y)\right]\longrightarrow \frac{i\omega \pi}{16r^2} +\mathcal{O}\left(\frac{1}{r^3} \right),
\end{equation}
\begin{equation}
S_3^{(2)}=\int_r^\infty\frac{dx}{x^3}\int_x^\infty q^2y\partial_y V_4^{(0)}(y)dy \longrightarrow \mathcal{O}\left(\frac{1}{r^3}\right),
\end{equation}
\begin{equation}
\begin{split}
&V_4^{(2)}=-\int_r^\infty\frac{x dx}{x^4-1}\int_1^x dy \left[2i\omega y\partial_y V_4^{(1)}(y)+i\omega V_4^{(1)}(y)+y \partial_y S_3^{(2)}(y)+S_3^{(2)}(y)\right]\\
&~~~~~~~~~\longrightarrow \frac{1}{96r^2}\left[-\pi^2 \omega^2+q^2\left(6\log 2-3\pi \right)\right]+\mathcal{O}\left(\frac{1}{r^3}\right).
\end{split}
\end{equation}
The boundary data $v_{_i}$ are straightforwardly read off from the above perturbative solutions.

\section*{Acknowledgements}
We would like to thank Alex Buchel, Christopher Herzog, Hong Liu, Giuseppe Policastro, Dam Son and Andrej Starinets for useful discussions related to this work. YB also benefited from discussions with Yun-Long Zhang and Xin Gao. YB would like to thank CRM, Universite de Montreal for the hospitality and financial support during the workshop ``Applications of AdS/CFT to QCD and condensed matter physics'', and Physics Department of the University of Connecticut where part of this work was done. This work was supported by the ISRAELI SCIENCE FOUNDATION grant \#87277111, BSF grant \#012124, the People Program (Marie Curie Actions) of the European Union's Seventh Framework under REA grant agreement \#318921; and the Council for Higher Education of Israel under the PBC Program of Fellowships for Outstanding Post-doctoral Researchers from China and India (2014-2015).

\providecommand{\href}[2]{#2}\begingroup\raggedright\endgroup

\end{document}